\documentclass[aps,prx,twocolumn,longbibliography,floatfix,superscriptaddress]{revtex4-1}

\usepackage{graphicx}
\usepackage{epstopdf}
\usepackage{epsfig}
\usepackage{amssymb,amsmath,stmaryrd,tabularx}
\usepackage{wrapfig}
\usepackage{bm}
\usepackage[T1]{fontenc}
\usepackage[center]{subfigure}
\usepackage[toc,page]{appendix}
\usepackage{float}
\usepackage{booktabs}
\usepackage{array}
\usepackage{blindtext}
\usepackage{amsthm}
\usepackage[dvipsnames]{xcolor}
\usepackage[normalem]{ulem}
\usepackage[percent]{overpic}
\usepackage{hyperref}

\begin{document}

\renewcommand{\L}{\mathcal L}
\newcommand{\F}{\mathcal F}
\renewcommand{\d}{\partial}
\renewcommand{\L}{\mathcal L}
\newcommand{\f}{{\tilde f ^{'}}}
\newcommand{\e}{\epsilon}
\renewcommand{\b}{\beta}
\newcommand{\w}{\omega}
\renewcommand{\a}{\alpha}
\newcommand{\avg}[1]{\left \langle #1 \right \rangle}
\newcommand{\uc}[1]{\left |\left | #1 \right | \right |_{\vec u}}
\newcommand{\ncr}[2]{\begin{pmatrix} #1 \\ #2 \\ \end{pmatrix}}
\let\vec\mathbf
\renewcommand{\[}{\begin{equation}}
\renewcommand{\]}{\end{equation}}

\newcommand{\A}{\mathcal A}
\newcommand{\B}{\mathcal B}
\renewcommand{\P}{\bar \psi}
\newcommand{\R}{\mathcal R}
\newcommand{\C}{\mathcal C}
\newcommand{\red}[1]{{\textcolor{red}{#1}}}
\definecolor{green}{rgb}{0,0.69,0.314}
\newcommand{\green}[1]{{\textcolor{green}{#1}}}

\newcommand{\fq}{E_{\vec q_n}^{ \vec u} (\vec r)}
\newcommand{\fqp}{E_{\vec q_{n'}}^{\vec u}(\vec r)} 
\newcommand{\fp}{E_{\vec p_n}^{ \vec u} (\vec r)}
\newcommand{\fpp}{E_{\vec p_{n'}}^{ \vec u} (\vec r)}
\newcommand{\ft}{E_{\vec t_n}^{ \vec u} (\vec r)}
\newcommand{\ftp}{E_{\vec t_{n'}}^{ \vec u} (\vec r)}

\newcommand{\kd}{\delta}

\newcommand{\n}{\tilde \rho}
\newcommand{\la}[1]{{\textcolor{red}{#1}}}

\title{Stress in ordered systems: Ginzburg-Landau type density field theory}

\author{Vidar Skogvoll}
\affiliation{PoreLab, The Njord Centre, Department of Physics, University of Oslo, P. O. Box 1048, 0316 Oslo, Norway}

\author{Audun Skaugen}
\affiliation{Computational Physics Laboratory, Tampere University, P.O. Box 692, FI-33014 Tampere, Finland}

\author{Luiza Angheluta}
\affiliation{PoreLab, The Njord Centre, Department of Physics, University of Oslo, P. O. Box 1048, 0316 Oslo, Norway}

\date{\today}

\begin{abstract}
We present a theoretical method for deriving the stress tensor and elastic response of ordered systems within a Ginzburg-Landau type density field theory in the linear regime. 
This is based on spatially coarse graining the microscopic stress which is determined by the variation of a free energy with respect to mass displacements.
We find simple expressions for the stress tensor for phase field crystal (PFC) models for different crystal symmetries in two and three dimensions. 
Using tetradic product sums of reciprocal lattice vectors, we calculate elastic constants and show that they are directly related to the symmetries of the reciprocal lattices. 
We also show that except for bcc lattices, there are regions of model parameters for which the elastic response is isotropic. 
The predicted elastic stress-strain curves are verified by numerical strain-controlled bulk and shear deformations. 
Since the method is independent of a reference state, it extends also to defected crystals. We exemplify this by considering an edge and screw dislocation in the simple cubic lattice.
\end{abstract}

\maketitle


\section{Introduction}

Classical deformation theories are formulated on the assumption that a solid is a deformable continuum medium on length scales much larger than the size of any microscopic structures. 
This macroscopic deformation field is independent of the system size such that scale invariance becomes a symmetry of the solid \cite{baggioliScaleInvariantSolids2020}. 
However, this property is lost for solids that are micron and submicron in size, which deform erratically while exhibiting an overall strain hardening with decreasing system size \cite{uchicSampleDimensionsInfluence2004,dimidukSizeaffectedSingleslipBehavior2005,greerPlasticitySmallsizedMetallic2011}. 
Small crystals are still sufficiently big in size compared to the atomic scale of their crystal lattice, such that the continuum approximation remains valid, and in fact, desirable for a theory that aims to describe macroscopic properties. 
While the elastic degrees of freedom can still be coarse grained to elastic fields, dislocations, which are the main carriers of low-temperature plastic slips, cannot be readily coarse grained due to their topological nature which induces long-range interactions and persistent correlations. 
Conventional plasticity theories assume that the representative volume element is sufficiently large to contain a statistically significant number of dislocations such that the plastic deformation can be described in terms of a single average quantity, e.g., the dislocation density tensor, while ignoring fluctuations around it. 
This coarse graining procedure breaks down at the micron and submicron scales because there are not sufficiently many dislocations to substantiate a continuum approximation for the dislocation density, and the correlation length becomes comparable to the crystal size. 
We are thus left to imagine that on these scales, dislocations remain discrete entities interacting through their internally generated stress fields.
Discrete dislocation dynamics models are formulated precisely on these premises and consider dislocations as mobile singularities in a linearly elastic medium, e.g., Ref. \cite{guruprasadSizeEffectsHomogeneous2008}. 
This modelling approach has been successful at reproducing qualitatively the scale-free statistical properties of plastic slip avalanches \cite{papanikolaouAvalanchesPlasticFlow2017,ovaskaExcitationSpectraCrystal2017, weissMildWildFluctuations2015} and the size dependence of plastic yield \cite{zhouDiscreteDislocationDynamics2010,huPredictingFlowStress2019}. 
The model is nonetheless empirical in the way reaction rates and dislocation mobilities are introduced as ad-hoc tuning parameters.

There are several field formulations which attempt to link atomic with continuum scales through hybrid continuum/discrete models \cite{limkumnerdMesoscaleTheoryGrains2006,salmanMinimalIntegerAutomaton2011,baggioLandautypeTheoryPlanar2019}, or by introducing free-tuning intrinsic length parameters as in strain-gradient plasticity theories~\cite{fleckStrainGradientPlasticity1994,liuMaterialLengthScale2017,lazarNonsingularStressStrain2005}. 
We are still lacking a theoretical model with no ad-hoc parameters that captures quantitatively the rich plastic behavior of small crystals while also being able to shed light on the microscopic mechanisms behind the macroscopic plastic instabilities and fluctuations. 
A promising contender is the phase-field crystal (PFC) model \cite{elderModelingElasticPlastic2004,elderPhasefieldCrystalModeling2007} which accommodates more naturally the linkage between atomic and continuum scales.
It models the crystal lattice as a continuous density field and encodes both the state of elastic deformation and the plastic slip. 
For this reason, it has been used to model various crystal-related phenomena \cite{emmerichPhasefieldcrystalModelsCondensed2012}. 
A caveat with the standard PFC model is that it lacks the separation of timescales between the overdamped dislocation motion and the very fast relaxation to equilibrium of elastic modes \cite{stefanovicPhasefieldCrystalsElastic2006,heinonenConsistentHydrodynamicsPhase2016}. 
Recently, we have proposed a way to remedy this by constraining the diffusive relaxation to accommodate instantaneous mechanical equilibrium on continuum scales \cite{skaugenSeparationElasticPlastic2018,salvalaglioCoarsegrainedPhasefieldCrystal2020}, which makes it possible to study how dislocations nucleate under stress \cite{skogvollDislocationNucleationPhasefield2021}.  

Our method of linking between the continuum scale of elasticity and the discrete nature of dislocations is based on computing the macroscopic stress tensor directly from the PFC free energy functional, hence the order parameter. 
We have done this derivation for a specific free energy in two dimensions in Ref.  \cite{skaugenDislocationDynamicsCrystal2018}. 
A generalization is needed to compute the stress field from an arbitrary free energy in any dimensions, and this we address in this paper.  
Density functional theories provide a similar conceptual technique for computing the microscopic stress from more ab-initio free energies and based on Irving-Kirkwood transport theory \cite{irving1950statistical, bartolottiConceptPressureDensity1980,dalcorsoDensityfunctionalTheoryMacroscopic1994,filippettiTheoryApplicationsStress2000,krugerStressesNonequilibriumFluids2017}. 
However, this stress is not coarse grained or maintained at mechanical equilibrium.  
In molecular dynamics models, the microscopic stress is also computed through the Irving-Kirkwood formula, a generalization of the virial expression of the equation of state to non-equilibrium systems \cite{tsaiVirialTheoremStress1979,lutskoStressElasticConstants1988}, but the system is confined to both atomic length scales and fast time scales. 
By contrast to these approaches, the PFC model with mechanical equilibrium handles multiples scales both in space and time.  
Another advantage of the PFC modelling formalism is that dislocations are emergent features, determined by the topological defects in the complex amplitudes obtained by the mode expansion of the crystal order parameter \cite{skaugenDislocationDynamicsCrystal2018}. 
We have shown that the profile of the macroscopic stress around a dislocation matches the analytical solutions from linear elasticity in the far-field and is regular at the dislocation core due to the smooth properties of the order parameter  
\cite{skaugenSeparationElasticPlastic2018,salvalaglioCoarsegrainedPhasefieldCrystal2020}. 
Thus, the formalism presented in this paper can be extended to plastic deformation and flow due to the presence of dislocations. 

In this paper, we propose a systematic method that links the macroscopic stress field which describes the deformation state of a continuum elastic medium with the microscopic stress field, which, in turn, is directly determined by the order parameter of the broken crystal symmetries (a crystal density field). 
The generic procedure is based on finding the microscopic stress through a variational calculus of an appropriate free energy with respect to mass displacements followed by a coarse graining procedure to upscale the microscopic stress to continuum scales. 
This method is valid for a Ginzburg-Landau type theory in which the free energy is given in terms of an order parameter and any order of its gradients. 
We provide several examples of free energies for crystals in two and three dimensions. 
Expanding the crystal order parameter in its reciprocal modes, we find that the elastic constants of the macroscopic stress are directly linked  to tetradic product sums (fourth order moment tensors) of the reciprocal lattices of the microscopic structure.  
In particular, this shows how the isotropic elastic response of the 2D hexagonal PFC arises directly from the six-fold symmetry of its reciprocal lattice, since only isotropic tetradic product sums can be formed from such vector sets \cite{chenMomentIsotropyDiscrete2011}.

The rest of the paper is structured as follows: 
In Sec. \ref{sec:microscopic_stress_fields}, we present the variational procedure for a microscopic one-body density and formally connect its expression with the chemical potential.
In Sec. \ref{sec:coarse_grained_stress}, we coarse grain the microscopic stress tensor over a representative volume element and show how it relates to the macroscopic stress in the linear regime. 
We then consider specific forms of the free energy in Sec. \ref{sec:PFC_models}, for which we derive explicit expressions for the stress tensor and compute the elastic constants. 
Finally, a brief summary and concluding remarks are given in Sec. \ref{sec:conclusion}.


\section{Microscopic stress fields}
\label{sec:microscopic_stress_fields}
The microscopic Cauchy stress $\tilde \sigma_{ij}$ can be determined by variational changes of a free energy $F$ with respect to adiabatic mass displacement variations $\delta \vec{x}$ through
\[
\delta F = - \int_{\Omega} d^D r \d_i \tilde \sigma_{ij} \delta x_j+\int_{\d \Omega} dS_i \tilde \sigma_{ij} \delta x_j,\label{eq:DivergenceOfStress}
\]
where $\d \Omega$ is the surface of the volume element $\Omega$ of dimension $D$.
In continuum mechanics, the stress is determined through a variation of the free energy with respect to an underlying displacement field $\vec u$, which determines how a medium has been deformed from some reference state. 
The stress definition of Eq. (\ref{eq:DivergenceOfStress}), however, is independent of such a reference state and we will show in Section \ref{sec:coarse_grained_stress} how we relate this definition to the continuum stress in the linear regime. 
In conventional density functional theory, $F$ is the sum of the ideal gas free energy $F_{id}[\tilde \rho]$,  an external potential energy $F_{ext}[\tilde \rho]$, and an excess free energy $F_{exc}$ which accounts for particle mutual interactions. The former two are expressed as functionals of the microscopic one-body density $\tilde \rho$ which is the ensemble-average of the density operator for $N$ particles
\[
\tilde \rho (\vec r) = \avg{\sum_{i=1}^N \delta(\vec r-\vec r_i)}_{\textrm{Ens}},
\]
while $F_{exc}$ must be approximated for practical purposes \cite{vrugtClassicalDynamicalDensity2020}. 
In this paper, we are interested in Ginzburg-Landau type field theories, in which $F_{exc}$ is expressed in terms of gradients, and its exact expression is typically determined by the symmetries of the ordered phase.
Thus, $F$ is given as a functional of $\tilde\rho$ and its gradients, $F[\tilde\rho] =\int d^D r \tilde f (\tilde\rho,\{\d_i \tilde\rho\}, \{ \d_{ij} \tilde \rho \},...)$, where $\tilde f$ is the free energy density. 
Therefore, variational changes in $F$ relate directly to variational changes in the microscopic density, and the corresponding conjugate variable defines the chemical potential
\[
\tilde\mu_c (\tilde \rho) = \frac{\delta F}{\delta \tilde\rho} = \frac{\partial \tilde f}{\partial \tilde\rho}-
\d_i
\frac{\d \tilde f}{\d (\d_i \tilde\rho)}
+
N(\{i,j\}) \d_{ij} \frac{\d \tilde f}{\d (\d_{ij} \tilde\rho)}\cdots
\]
where $N(\{i,j\}) = (1+\delta_{ij})/2$ is a necessary prefactor to not over-count contributions from the off-diagonal variables (see Appendix \ref{sec:appendix_combinatorial_factor}).

To derive  Eq. (\ref{eq:DivergenceOfStress}) for variational changes of $F$ in terms of $\delta\vec x$, we use that mass density is a locally conserved quantity, so that its variations $\delta\tilde \rho$ are determined by the mass displacement variations through the conservation law, written as 
\[
\delta \tilde\rho = -\d_j (\tilde\rho \delta x_j). 
\label{eq:delta_n_function_of_u}
\]
This implies that the variational of $F$ relates to $\delta \vec x$ as  
\[\begin{split}
\delta F &= 
\int_{\Omega} d^D r \tilde \mu_c \delta \tilde \rho \\
&=
\int_{\Omega}d^D r (\tilde\rho \d_j \tilde \mu_c) \delta x_j-\int_{\d \Omega} dS_j (\tilde\mu_c \tilde\rho) \delta x_j.
\end{split}\]
Identifying this with the expression in Eq. (\ref{eq:DivergenceOfStress}), we obtain that a net mechanical force leads to mass transport along the chemical potential gradient, namely
\begin{equation}
\d_i \tilde \sigma_{ij} = -\n \d_j \tilde\mu_c.
\label{eq:definition_of_sigma_ij_in_temrms_of_muc}
\end{equation}
This expression tells us equivalently that when the system is in chemical equilibrium (steady-state microscopic density) then the associated microscopic stress is in mechanical equilibrium and vice versa. 
To obtain an explicit expression for $\tilde \sigma_{ij}$, we consider a free energy density $\tilde f(\tilde \rho, \{\d_i \tilde \rho\},\{\d_{ij} \tilde \rho\})$ that only depends on $\tilde \rho$ and its first and second order gradients, so that the free energy changes by
\[\begin{split}
\delta F &= 
\int_{\Omega} d^D r 
\left (\frac{\d \tilde f}{\d \tilde \rho} \delta \tilde \rho 
+\frac{\d \tilde f}{\d (\d_i \tilde \rho)} \delta (\d_i\tilde \rho) \right . \\ 
&\hspace{3cm}
\left .
+N(\{i,j\}) 
\frac{\d \tilde f}{\d (\d_{ij} \tilde \rho)} \delta (\d_{ij} \tilde \rho) \right ).
\end{split}\]
Using Eq. (\ref{eq:delta_n_function_of_u}), and repeated integration by parts and renaming of indices, we obtain that (up to some surface terms)
\[\begin{split}
\delta F 
&= \int_{\Omega} d^D r \delta x_j \d_i \left [ 
(\tilde f - \tilde \mu_c \tilde \rho) \delta_{ij}
-\frac{\d \tilde f}{\d (\d_i \tilde \rho)} \d_j \tilde \rho
\right . \\ 
&
\hspace{-.7cm}
\left .
+N(\{i,m\})\left (
\d_m \frac{\d \tilde f}{\d (\d_{im} \tilde \rho)} 
\right )
\d_j \tilde \rho
-N(\{i,m\})\frac{\d \tilde f}{\d (\d_{im} \tilde \rho)} \d_{mj} \tilde \rho
\right ]\\
&\equiv
\int_\Omega d^D r \delta x_j \d_i \tilde \sigma_{ij},
\label{eq:stress_tensor_from_variational_calculus}
\end{split}\]
with the microscopic stress tensor defined as
\[
\tilde \sigma_{ij} 
= (\tilde f-\tilde\mu_c \n) \delta_{ij}
+\tilde h_{ij},
 \label{eq:general_stress_tensor}
\]
where $\tilde h_{ij}$ arises from the gradient expansion of the non-local interaction and is given by
\[
\tilde h_{ij} = -\f_i \d_j \tilde \rho - \f_{im} \d_{jm} \tilde \rho + (\d_m \f_{im})\d_j \tilde \rho.
\]
Here, we have introduced the notation $\f_i = \d \tilde f/\d(\d_i \tilde \rho)$ and $\f_{im} = N(\{i,m\}) \d \tilde f / \d(\d_{im} \tilde \rho)$.
Taking the divergence of Eq. (\ref{eq:general_stress_tensor}), a lengthy, but straight-forward calculation shows that it satisfies the force balance Eq. (\ref{eq:definition_of_sigma_ij_in_temrms_of_muc}) (see the general derivation in Appendix \ref{sec:appendix_proof_of_generalization}).

Closer inspection of the method outlined here reveals a gauge freedom in the determination of the microscopic stress. 
Both the force balance Eq. (\ref{eq:definition_of_sigma_ij_in_temrms_of_muc}) and the variation calculus of Eq. (\ref{eq:stress_tensor_from_variational_calculus}) only define the stress tensor up to a divergence free contribution. 
Additionally, since the dynamics is independent of any constant surface contribution to the free energy, local free energies are undetermined up to a divergence in $\tilde f$. 
The ambiguity is of no physical importance since it does not change the force density and is fundamentally associated with the difficulty in attributing a local (point-wise) energy contribution to a system in which there are non-local interactions
\cite{nielsenStressesSemiconductorsInitio1985,chettyFirstprinciplesEnergyDensity1992,filippettiTheoryApplicationsStress2000}.
Indeed, for structures with an intrinsic length scale, such as crystals, coarse graining over a representative volume will in part remove this ambiguity, as we will demonstrate shortly.  

Equation (\ref{eq:general_stress_tensor}) suggests a generalization which is valid for a free energy density $\tilde f (\tilde\rho,\{\d_i \tilde\rho\}, \{ \d_{ij} \tilde \rho \},...)$ that is a function of arbitrary density gradients, where $\tilde h_{ij}$ is replaced by
\[
\tilde h_{ij} = \sum_{\a=1}^\infty \tilde M_{ij}^{(\a)},
\]
where
\[
\tilde M^{(\a)}_{ij}
=
 \sum_{\b=1}^{\a}
 (-1)^\b
 (\d_{m_1 ... m_{\beta-1}} \f_{m_1...m_{\a-1}i})
 \d_{j m_\b ... m_{\a-1}} \n,
 \label{eq:Mij_alpha}
\]
and the short-form notation has been generalized to
\[
\f_{m_1 ... m_{\a}}
=
N(\{m_i\})
\frac{\d \tilde f}{\d (\d_{m_1...m_\a} \n)}
\label{eq:f_notation}.
\]
The combinatorial factor is the inverse of the multinomial coefficient
\[
N(\{m_i\}_{i=1}^\a)
=
\frac{N_x! N_y! N_z!}{\a!}
\]
where $N_x$, $N_y$ and $N_z$ are the numbers of elements in $\{m_i\}_{i=1}^\a = \{m_1,...,m_\a\}$ that equal $x$, $y$ and $z$, respectively. 
While it is possible to redo the variational calculus for an arbitrary number of gradients, the easiest is to confirm that the generalization satisfies the force balance Eq. (\ref{eq:definition_of_sigma_ij_in_temrms_of_muc}). 
In Appendix \ref{sec:appendix_combinatorial_factor}, we show that $\d_i \tilde h_{ij} =  \tilde \mu_c \d_j \tilde \rho - \d_j \tilde f$ from which Eq. (\ref{eq:definition_of_sigma_ij_in_temrms_of_muc}) follows.  

As an example of this general expression of the microscopic stress, we take the Ginzburg-Landau free energy $\tilde f(\n,\{\d_i \n\} )$, for which Eq. (\ref{eq:general_stress_tensor}) reduces to
\[
\tilde \sigma_{ij} = (\tilde f-\tilde\mu_c \n)\delta_{ij} - \frac{\partial \tilde f}{\partial (\partial_i \n)} \d_j \n 
\]
which is the expression derived in Ref. \cite{krugerStressesNonequilibriumFluids2017}.
For a free energy density $\tilde f(\tilde \rho,\{\d_i \tilde \rho\},\{\d_{ij} \tilde \rho\})$ dependent on second order gradients, such as the basic Swift-Hohenberg free energy functional used in the PFC model, we get an expression of Eq. (\ref{eq:general_stress_tensor}) which is the general form of the stress tensor used in Ref. \cite{skaugenDislocationDynamicsCrystal2018}. 
It should be noted that the stress tensor in Ref. \cite{skaugenDislocationDynamicsCrystal2018} omits the combinatorial factor $N(\{i,j\})$ since this included only terms diagonal in $ij$, for which $N(\{i,j\}) = 1$. 
It also lacks the second term $-\tilde \mu_c \n$ in the isotropic part of the stress tensor as this arises from considering mass-conserving deformations, which were not considered in Ref. \cite{skaugenDislocationDynamicsCrystal2018}.
For more general free energy expressions, e.g., those given in Refs. \cite{lifshitzTheoreticalModelFaraday1997,wuPhasefieldcrystalModelFcc2010,wuControllingCrystalSymmetries2010,mkhontaExploringComplexWorld2013}, the general expression must be employed and we present in this article some of these expressions.


\section{Coarse grained description: Continuum limit}
\label{sec:coarse_grained_stress}

The notion of the stress tensor defined in the previous section is valid for any density field $\n$.
A crystal as a continuum elastic medium, by contrast, has far fewer degrees of freedom and is typically characterized by a \textit{macroscopic} density field  $\rho(\vec r) = \avg {\tilde \rho}(\vec r)$, defined as a spatial average of $\tilde \rho$ over a unit representative volume, which for a crystal is given by the lattice unit $a_0$.
For the remainder of this paper, we consider the Gaussian convolution
\[
\avg{\tilde \rho}(\vec r) = \int d^D r'\frac{\tilde \rho(\vec r')}{(2\pi a_0^2)^{D/2}} \exp\left (-\frac{(\vec r-\vec r')^2}{2a_0^2}\right)
\label{eq:coarse_graining_definition}
\]
as the definition of a coarse graining procedure.
The evolution of this density is dictated by minimizing the coarse grained free energy density $f = \langle \tilde f\rangle$. 
The deformation varies on scales much larger than the crystal lattice, so it is described by a macroscopic (slowly varying) displacement field $\vec u$. This appears as changes in the phases of the complex amplitudes of the demodulated density field. 
In linear elasticity, the strain field $e_{ij}$ is the symmetric part of the displacement gradient, $e_{ij} = \d_{(i}u_{j)}$, where  $[\cdots]_{(ij)} = ([\cdots]_{ij} + [\cdots]_{ji})/2$ is the symmetrization over indices $ij$.
The constitutive law connecting this strain with the stress emerges on continuum scale, after coarse graining the atomic-scale interactions.
The Eulerian picture  \cite{chaikinPrinciplesCondensedMatter1995} provides the natural framework to define mass displacement variations $\delta \vec x$ and how they induce changes in $\vec u$ and $\rho$.
Namely, the volume element and macroscopic density change according to $d^D r' \rightarrow (1+\d_k \delta x_k) d^D r'$ and $\rho \rightarrow (1-\d_k \delta x_k) \rho$. Additionally, the distance between planes of constant phase changes by $\delta \vec x$, so that the linear strain tensor transforms as $e_{ij} \rightarrow e_{ij} + \d_{(i} \delta x_{j)}$.
Thus the variation in free energy becomes 
\[\begin{split}
\delta F &= \int_{\Omega}
d^D r\left (
f \d_k \delta x_k  - \rho \left . \frac{\d f}{\d  \rho}  \right |_{e_{ij}} \d_k \delta x_k + \left . \frac{\d f}{\d e_{ij}} \right |_{\rho} \d_i \delta x_j \right )\\
&=
\int_{\d \Omega} dS_i \sigma_{ij} \delta x_j
-
\int_{\Omega}  \d_i \sigma_{ij} \delta x_j
\label{eq:DefinitionOfStressThroughIntegral}
\end{split}\]
where the macroscopic stress tensor $\sigma_{ij}$ is given by
\[
\sigma_{ij} = \left (  f - \rho\left .\frac{\d  f}{\d \rho}\right |_{e_{ij}} \right )\delta_{ij} + \left . \frac{\d f}{\d e_{ij}} \right |_{\rho}.
\label{eq:crystal_stress_tensor}
\]

\begin{figure}
    \centering
    \includegraphics[scale=0.9]{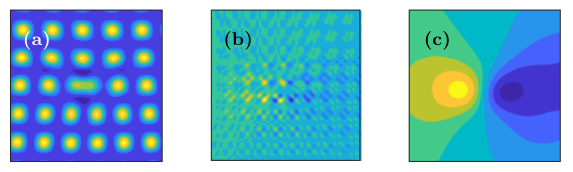}
    \caption{(a) microscopic density field $\tilde \rho$ in a 2D PFC system with square lattice symmetry and an edge dislocation, (b) its microscopic stress field $\tilde \sigma_{xy}$, and (c) the macroscopic stress field $\sigma_{xy} = \avg{\tilde \sigma_{xy}}$, the object of continuum deformation theories. 
    }
    \label{fig:coarse_grain_illustration}
\end{figure}

Limiting our attention to crystals, we are interested in a macroscopic stress tensor in the form given by Eq. (\ref{eq:crystal_stress_tensor}). 
Assuming that the microscopic density field $\n$ that minimizes the free energy is given by a Bravais lattice with lattice constant $a_0$, we can coarse grain Eq. (\ref{eq:general_stress_tensor}) with respect to this length scale.
In equilibrium, $\tilde \mu_c$ is spatially constant, so for a small deviation from equilibrium, such as given by a macroscopically varying density field $\delta \rho(\vec r)$ or a macroscopic displacement field $\vec u$, $\tilde \mu_c$ is slowly varying and invariant under coarse graining, $\mu_c = \avg{\tilde \mu_c} = \tilde \mu_c$. 
Thus, $\avg{\tilde \mu_c \n} = \tilde \mu_c \avg{\n} = \mu_c \rho$ and by coarse graining Eq. (\ref{eq:general_stress_tensor}), we find
\[
\avg {\tilde \sigma_{ij}}
=
(f - \mu_c \rho) \delta_{ij} + h_{ij},
\label{eq:coarse_grained_general_stress}
\]
where $h_{ij} = \avg{\tilde h_{ij}}$.
One can show that $\tilde \mu_c = \delta F/\delta \n = \delta F/\delta \rho$ which allows us to identify $\sigma_{ij} =\avg {\tilde \sigma_{ij}} $ if 
\[
\left . \frac{\d f}{\d e_{ij}} \right )_{\rho} = h_{ij},
\label{eq:hij_as_derivative_of_strain}
\]
which shows that $h_{ij}$ is the thermodynamic conjugate of the strain at constant macroscopic density in the linear regime.
Figure \ref{fig:coarse_grain_illustration} shows an example of a microscopic density field, its associated microscopic stress and the macroscopic stress field after coarse graining. 
While the microscopic stress tensor describes internal stresses across all length scales, the macroscopic stress tensor $\sigma_{ij}$ describes stresses between representative volume elements bigger than that of the unit cell. 
Thus, while $\d_i \tilde \sigma_{ij}=0$ (complete chemical equilibrium, Eq. (\ref{eq:definition_of_sigma_ij_in_temrms_of_muc})) implies $\d_i \sigma_{ij} = 0$ the converse is not true in general. 
In fact, it is known that for dislocation dynamics, the evolution of long wavelength distortions (macroscopic disturbances) is much faster than the diffusive dynamics of local distortions, such as the motion of dislocation lines.
The typical dissipative evolution of dynamical density functional theory and phase-field modelling makes no explicit distinction between the evolution of disturbances at short and long wavelengths, which has led to the development of several theories that treat this separation of time-scales explicitly \cite{heinonenPhasefieldcrystalModelsMechanical2014,heinonenConsistentHydrodynamicsPhase2016,skaugenSeparationElasticPlastic2018}.

Since the gauge invariance of the stress tensor is related to the difficulty of having a well-defined local free energy under interactions, the act of coarse graining with respect to a length scale defined by the density field periodicity partly removes this ambiguity.
As an example, consider the product $\avg{(\nabla A)B}$ under coarse graining of two fields $A,B$ that vary on the microscopic scale respective of the underlying lattice, but are slowly varying under coarse graining. 
Such fields can be expanded in terms of slowly varying amplitudes $A_n(\vec r),B_n(\vec r)$ as $A=\sum_{\vec K \in \R} A_n(\vec r)e^{i\vec K \cdot \vec r}$, $B=\sum_{\vec K \in \R} B_n(\vec r)e^{i\vec K \cdot \vec r}$, where $\R$ is the reciprocal lattice of the microscopic structure.
We find
\[\begin{split}
\avg{(\nabla A) B}
&=
\sum_{\vec K_n\in \R}\sum_{\vec K_{n'} \in \R} 
\avg{\nabla (A_n(\vec r)e^{i\vec K_n \cdot \vec r}) B_n(\vec r) e^{i\vec K_{n'}\cdot \vec r}}\\
&\approx
\sum_{\vec K_n\in \R}\sum_{\vec K_{n'} \in \R} 
\left [
(\nabla A_n(\vec r) + i \vec K_n A_n) B_n(\vec r)
\right .\\ 
&\hspace{4cm}\left .
\times\avg{
e^{i (\vec K_n  + \vec K_{n'}) \cdot \vec r}
}\right ],
\label{eq:avg_nabla_A_B}
\end{split}\]
where we have used that the fields $A_n(\vec r),B_n(\vec r)$ vary slowly on the periodicity of the lattice to take the coarse graining through. 
The coarse grained value of $e^{i (\vec K_n  + \vec K_{n'}) \cdot \vec r}$ will only be non-zero at resonance, given by $\vec K_{n'} = -\vec K_n$, i.e. $\avg{
e^{i (\vec K_n  + \vec K_{n'}) \cdot \vec r}
} \equiv \delta_{n',-n}$ \cite{footnote_coarse_grained_delta_function}, and using that for slowly varying amplitudes $|\nabla A_n(\vec r)|\ll |\vec K A_n  (\vec r)|$, we find 
\[\begin{split}
\avg{(\nabla A) B}
&\approx
\sum_{\vec K_n\in \R}
i \vec K_n A_n (\vec r) B_{-n}(\vec r) \\
&=
\sum_{\vec K_n\in \R}
i (-\vec K_{-n})  B_{-n}(\vec r) A_n(\vec r)
\approx
-\avg{A (\nabla B)},
\label{eq:gradient_shfiting_under_coarse_graining}
\end{split}\]
as can be shown by expanding the right-hand side under similar assumptions.
The exact difference between the left- and right-hand side of this equation is given by 
\[
\avg{(\nabla A) B}-(-\avg{A (\nabla B)})
=
\avg{\nabla(AB)}
=
\nabla \avg{AB},
\]
since the gradient operator commutes with the coarse graining operation as can be seen by 
\[\begin{split}
\d_i \avg{\tilde X}
&=
\int d^D r'\frac{\tilde X(\vec r')}{(2\pi a_0)^{D/2}}\d_i \exp\left (-\frac{(\vec r-\vec r')^2}{2a_0^2}\right)\\
&=
-\int d^D r'\frac{\tilde X(\vec r')}{(2\pi a_0)^{D/2}}\d_{i'} \exp\left (-\frac{(\vec r-\vec r')^2}{2a_0^2}\right) \\
&=
\int d^D r' \frac{ \d_{i'} \tilde X(\vec r')}{(2\pi a_0)^{D/2}} \exp\left (-\frac{(\vec r-\vec r')^2}{2a_0^2}\right) 
= \avg{\d_i \tilde X}.
\end{split}\]
Thus, by employing Eq. (\ref{eq:gradient_shfiting_under_coarse_graining}) to rewrite expressions, we are ignoring the variations in the coarse grained boundary terms. 
In the case of a crystalline lattice, we have seen computationally that this identity holds far beyond the regime of linear elasticity. 
Equation (\ref{eq:gradient_shfiting_under_coarse_graining}) shows how gradient terms of the microscopic stress tensor can be rewritten under coarse graining, indicating that different description of the stress on the microscopic scale are equivalent upon coarse graining. 
In particular, it allows rewriting Eq. (\ref{eq:Mij_alpha}) in a coarse grained form as 
\[
\avg{
\tilde M_{ij}^{(\alpha)}
}
\approx
- \a
\avg{
\f_{m_1...m_{\a-1} i} \d_{jm_1...m_{\a-1}} \tilde \rho
}.
\label{eq:Mijalpha_simplified_by_partial_integration}
\]
As will be shown for the PFC models introduced in the next section, this expression will be symmetric in the indices, $i\leftrightarrow j$, indicating that coarse graining will make the stress tensor explicitly symmetric.
This is consistent with result of recent work which used MD simulations to show that the symmetric nature of stress tensor is intimately linked with the continuum assumption and may break down on a microscopic resolution \cite{rigelesaiyinAsymmetryAtomiclevelStress2018}.


\section{Application to phase-field crystal models}
\label{sec:PFC_models}

\begin{figure}
    \centering
    \includegraphics[scale=0.95]{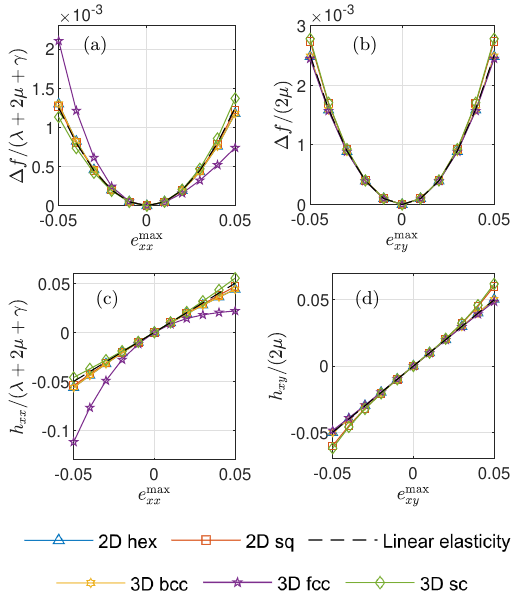}
    \caption{Difference in the free energy density
    $\Delta f = \avg{\tilde f(\psi')} - \avg{\tilde f(\psi^{eq})}$ in the center region (red dots in Figs. \ref{fig:bulk_and_shear_deformation}(c-d)) for $\psi'(\vec r)$, Eq. (\ref{eq:deformation_procedure}), strained by (a) the bulk displacement field $\vec u^{(B)}$, Eq. (\ref{eq:bulk_displacement_field}), and (b) the shear displacement field $\vec u^{(S)}$, Eq. (\ref{eq:shear_displacement_field}).
    The non-zero stress components for the PFC strained by the (c) bulk and (d) shear displacement fields.
    In all plots, the dashed lines indicate the prediction from linear elasticity, Eqs. (\ref{eq:hex_bulk_elastic_energy}-\ref{eq:hex_shear_elastic_energy}).}
    \label{fig:energy_and_stress_of_bulk_and_shear_displacement}
\end{figure}

In this section, we consider forms of the free energy $F$ specific to systems with different crystal symmetries.
An already well-established minimal model for this is the PFC which was introduced as a phenomenological field theory to model crystallization and related phenomena \cite{elderModelingElasticPlastic2004}.
We investigate five PFC models: 2D hexagonal (2D hex), 2D square (2D sq), 3D bcc, 3D fcc and 3D simple cubic (3D sc), using established free energy functionals for the first four and an adapted PFC model for the sc phase. 
As customary for phase-field modelling, we employ the notation $\psi$ for the microscopically varying density field under consideration.
The stress tensor $\tilde\sigma_{ij}$ is defined in terms of a microscopic density field $\n$, so its expression in terms of $\psi$ depends on the exact connection between these two quantities. 
Here, we define $\psi \equiv \tilde\rho$, and consider the elastic response of $\psi$ during an adiabatic deformation at constant macroscopic density $\avg \psi$, which is achieved by the following transformation of the field
\[
\psi'(\vec r) = \psi^{eq}(\vec r - \vec u),
\label{eq:deformation_procedure}
\]
where $\psi^{eq}(\vec r)$ is the unstrained equilibrium crystal configuration and $\vec u$ is an arbitrary macroscopic displacement field.
We employ the method developed in the previous sections to find explicit forms of the stress tensors in terms of $
\psi$. 
Then, for each PFC model, we expand the ground state in an appropriate number of reciprocal lattice modes to obtain expressions for the elastic constants in terms of the reciprocal mode amplitudes by use of tetradic product sums. 
For the established free energy functionals, we find the elastic constants in agreement with previous works.
From the expressions of the elastic constants, we find that there are particular model parameters for which all the PFC models exhibit isotropic elasticity except for the 3D bcc model which always exhibits anisotropic elastic behavior. 
The results are summarized in Table \ref{tab:elasticity_summary}. 
\begin{table*}[]
    \centering
    \begin{tabular}{lllllll}
    \toprule
    PFC model & $h_{ij}$  & Elastic constants:& Isotropic elastic domain\\
    \midrule
    2D hex & $-2\avg{(\L_1 \psi) \d_{ij} \psi}$ & $\lambda = 3A^2$&  Always \\ 
    &&$\mu = 3A^2 $ \\
    &&$\gamma = 0 $ \\
    \midrule
    2D sq & $-2\avg{(\L_1 \L_2 \psi)(\L_1 + \L_2) \d_{ij} \psi}$& $\gamma = 16B^2$ & $r=-\frac{25}{3}\P^2$ \\
    &&$\mu = 16B^2 $ \\
    &&$\gamma = 8A^2-32B^2  $ \\
    \midrule
    3D bcc & $-2\avg{(\L_1 \psi) \d_{ij} \psi}$ & $\lambda = 4 A^2$  & Never\\
    &&$\mu = 4 A^2 $ \\
    &&$\gamma = -4A^2 $ \\
    \midrule
    3D fcc & $-2\avg{(\L_1 \L_{\frac 4 3} \psi)(\L_1 + \L_{\frac 4 3}) \d_{ij} \psi}$&$\lambda = \frac{32}{81} A^2$ &  $r = -\frac{255 }{49}\P^2$ \\
    &&$\mu = \frac{32}{81} A^2 $ \\
    &&$\gamma = \frac{32}{81} (2B^2 - A^2) $ \\
    \midrule
    3D sc & $-2\avg{(\L_1 \L_2 \L_3 \psi)(\L_2 \L_3 + \L_1 \L_3 + \L_1 \L_2) \d_{ij} \psi}$ & $\lambda = 16 B^2 + 128 C^2$   
    & $r \approx -2.10144 \P^2$\\
    &&$\mu = 16 B^2 + 128 C^2 $ \\
    &&$\gamma = 32 A^2 - 16 B^2 - 256 C^2 $ \\
    \bottomrule
    \end{tabular}
    \caption{$h_{ij}$ and its associated elastic constants in terms of amplitudes ($A,B,C$) of the mode expansion for different PFC models. 
    The free energy functionals for the different PFC models are given by Eqs. (\ref{eq:free_energy_dens_hex_PFC}), (\ref{eq:PFC_square_free_energy}), (\ref{eq:PFC_bcc_free_energy}), (\ref{eq:PFC_fcc_free_energy}), and (\ref{eq:SC_free_energy_density}), respectively, where $\L_X = X+\nabla^2$. 
    The elastic constants can be expressed in Voigt notation by $C_{11} = \lambda+2\mu+\gamma$, $C_{12} = \lambda$, $C_{44}=\mu$. 
    The last column gives the relationship between PFC model parameters $r,\bar \psi$ for which the elastic response is isotropic.}
    \label{tab:elasticity_summary}
\end{table*}
We then prepare a $60\times60$ ($\times 5$) unit cells periodic lattice for the 2D (3D) PFC models and numerically subject them to two types of distortions and calculate the elastic response and stress given by $h_{ij}$. 
The results are shown in Fig. \ref{fig:energy_and_stress_of_bulk_and_shear_displacement} and demonstrate that $h_{ij}$ measures the energetic response for a deformation at constant macroscopic density, as suggested by Eq. (\ref{eq:hij_as_derivative_of_strain}). 

The field transformation of Eq. (\ref{eq:deformation_procedure}) is a strain-controlled adiabatic deformation at constant macroscopic density $\bar \psi$, with no diffusive relaxation. 
We will thus recover the elastic constant tensor $C_{ijkl}$ at constant macroscopic density. 
These are not the same elastic constants as will be obtained if $\rho$ and $\vec u$ are not varied independently, and differ, for instance, from the elastic constants at constant chemical potential or at constant vacancy concentration \cite{chaikinPrinciplesCondensedMatter1995}.
Denoting the free energy under strain $e_{ij}$ as $F_{e_{ij}}$, we are thus focusing on strains in which only the planes of constant phase are displaced, using the equilibrium amplitudes $\{A_{\vec K}^{(0)}\}$ of the reciprocal modes $\{\vec K\}$.
This is equivalent to straining the reciprocal lattice vectors $\vec K \rightarrow \vec K_{e_{ij}}$, so that $F_{e_{ij}} = F[A_{\vec K}^{(0)},\{\vec K_{e_{ij}} \},\bar \psi, V]$.
If the order parameter is interpreted as a one-body density, the physical process of straining at constant macroscopic density requires counter-acting vacancy diffusion, unless the applied strain is traceless. 
The deformation is adiabatic in the sense that no minimization of the free energy at the given strain is performed. 
Ref. \cite{huterNonlinearElasticEffects2016} considered isothermal strain-controlled deformation at constant macroscopic density by also minimizing the free energy under strain, given by $F_{e_{i j}} = \min_{A_{\vec K}} F[\{A_{\vec K}\},\{\vec K_{e_{i j}} \},\bar \psi, V]$.
Since the equilibrium values of the amplitudes are minima in configuration space by definition, this deformation will lead to non-linear effects, and does not influence the elastic constants. 
Refs. \cite{wangElasticConstantsStressed2018,ainsworthPhaseFieldCrystal2019} considered strain-controlled isothermal deformation including the resulting volumetric deformation of the macroscopic density $\bar \psi \rightarrow \bar \psi_{e_{ij}}$ as well as induced changes in the region volume $V\rightarrow V_{e_{ij}}$, i.e.,  $F_{e_{ij}} = \min_{A_{\vec K}} F[\{A_K\},\{\vec K_{e_{i j}} \},\bar \psi_{e_{ij}}, V_{e_{ij}}]$.
It showed that the elastic constants associated to such a deformation differ from those obtained here or in previous works, and are also dependent on the exact connection between the order parameter $\psi$ and the physical one-body density.
This type of deformation corresponds to a mass displacement at constant vacancy concentration and hence these elastic constants could be derived from the stress tensor of Eq. (\ref{eq:coarse_grained_general_stress}).
However, in this case, the variational procedure must also be reevaluated under the specific connection between the order parameter and the physical one-body density.
In Ref. \cite{skogvollDislocationNucleationPhasefield2021}, we performed stress-controlled isothermal and quasi-static deformation of the PFC for which we found the nucleation of crystal defects occurring at strains $ |e_{ij} |\approx 0.1$. 


\subsection{2D Hexagonal lattice}

\begin{figure}
    \centering
    \includegraphics[width=0.4\textwidth]{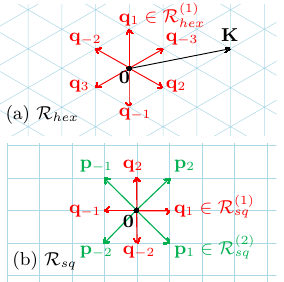}
    \caption{
    Reciprocal lattices of the 2D crystal structures.
    (a) Modes $\R_{hex}$ of the hexagonal reciprocal lattice where $\vec K$ denotes a general mode: $\R_{hex}^{(1)} = \{\vec q_{n} \}$ are the closest reciprocal lattice modes ($\vec q_n^2 = 1$). (b) Modes $\R_{sq}$ of the square reciprocal lattice: $\R_{sq}^{(1)} = \{\vec q_{n} \}$ are the closest reciprocal lattice modes ($\vec q_n^2 = 1$) and $\R_{sq}^{(2)} = \{\vec p_n\}$ are the next-to-closest modes ($\vec p_n^2 = 2$). 
    }
    \label{fig:2D_reciprocal_lattices}
\end{figure}

In its simplest form, the PFC free energy is based on the Swift-Hohenberg free energy given by $F=\int d^2 r \tilde f^{(hex)}$, with the free energy density
\[
\tilde f^{(hex)} = \frac{1}{2} (\L_1\psi)^2 + \frac{r}{2} \psi^2 +\frac{1}{4} \psi^4,
\label{eq:free_energy_dens_hex_PFC}
\]
where $\L_1 = 1+\nabla^2$ and $r$ is a parameter which is proportional to the deviation from the critical temperature. 
For the free energy density given in Eq. (\ref{eq:free_energy_dens_hex_PFC}), we find 
\[
\begin{split}
\f_{m_1} &= 0 \\
\f_{m_1 m_2} &= \L_1 \psi \delta_{m_1 m_2},\\
\end{split}
\] 
which gives
\[
\tilde h_{ij} =
- (\L_1 \psi) \d_{ij} \psi
+ (\d_i \L_1 \psi) \d_j \psi.
\label{eq:micro_stress_hex_PFC}
\]
For $r<0$ and a range of parameters $\P$, the free energy $F$ is minimized in two dimensions by a hexagonal lattice with lattice constant $a_0 = 4\pi/\sqrt 3$. 
Thus, for a perfect lattice, the density field $\psi$ can be expressed as a superposition of periodic modes in the reciprocal space associated to that lattice,
\[
\psi^{eq}_{hex} (\vec r) = \P + \sum_{\vec K \in \R_{hex} \setminus \{\vec 0\}} A_{\vec K}  e^{i\vec K \cdot \vec n},
\label{eq:equilibrium_psi_in_reciprocal_modes}
\]
where $\vec K$ is a non-zero mode of the hexagonal reciprocal lattice $\R_{hex}$, see Fig. \ref{fig:2D_reciprocal_lattices}(a), and $A_{\vec K}$ is the corresponding amplitude. 
When $|r| < 1$, the equilibrium state $\psi^{eq}_{hex}(\vec r)$ is well approximated in the one-mode expansion in terms of the principal reciprocal lattice vectors
\[
\psi_{hex}^{eq} (\vec r) \approx \P + A_{hex} \sum_{\vec q_n \in \R_{hex}^{(1)}} e^{i\vec q_n \cdot \vec r},
\label{eq:mode_approximation_hex_PFC}
\]
where $\R_{hex}^{(1)} = \{\vec q_{-3},\vec q_{-2},\vec q_{-1},\vec q_1,\vec q_2,\vec q_3\}$ are the closest non-zero modes on the hexagonal reciprocal lattice, which can be chosen as
\[
\begin{array}{l}
    \vec q_1 = (0,1)  \\
    \vec q_2 = (\sqrt 3/2,-1/2) \\
    \vec q_3 = (-\sqrt 3/2,-1/2),
\end{array}
\label{eq:Rhex1}
\]
$\vec q_{-n} = -\vec q_n$ and $A_{hex}$ is the equilibrium amplitude.
$A_{hex}$ is determined by inserting $\psi_{eq}^{hex}$ into the free energy density of Eq. (\ref{eq:free_energy_dens_hex_PFC}), averaging over a unit cell and minimizing with respect to $A_{hex}$ \cite{elderModelingElasticPlastic2004}.
Given the length scale of the lattice constant, we can define the stress tensor (associated to a continuum elastic medium) in terms of $\psi$ by coarse graining 
\[
\sigma_{ij}^{(2D~hex)}
=
\delta_{ij} (f-  \mu_c  \avg{\psi} ) + h_{ij}^{(2D~hex)},
\]
where $f = \avg {\tilde f^{(hex)}}$, $\mu_c = \delta F/\delta \psi$, and we have used Eq. (\ref{eq:Mijalpha_simplified_by_partial_integration}) to write 
\[
h_{ij}^{(2D~hex)} = - 2\avg{(\L_1 \psi) \d_{ij} \psi}.
\label{eq:hij_hex_PFC}
\]
The elastic coefficients $C_{ijkl}$ of the corresponding hexagonal lattice can be computed by deforming the one-mode approximation by a macroscopic displacement field $\vec u$ according to the field transformation of Eq. (\ref{eq:deformation_procedure}), which gives to first order in the distortion $\d_k u_l$ (see Appendix \ref{sec:appendix_hexagonal_mode_expansion})
\[
h_{ij}^{(2D~hex)} = 4 A_{hex}^2 \d_k u_l \sum_{\vec q_n \in \R_{hex}^{(1)}} q_{ni} q_{nj} q_{nk} q_{nl},
\]
where $q_{ni}$ is the $i$th Cartesian coordinate of the reciprocal lattice vector $\vec q_n$. 
This shows that the elastic constant are directly determined by the tetradic product sum of $\R_{hex}^{(1)}$.
This is a general feature of all the PFC models which we consider in this paper. 
It is given by
\[
\sum_{\vec q_n \in \R_{hex}^{(1)}} q_{ni} q_{nj} q_{nk} q_{nl} = \frac{3}{4} (\delta_{ij} \delta_{kl} + 2 \delta_{k(i} \delta_{j)l}),
\]
as can be shown by checking all components.
Thus, we find
\[
C_{ijkl} = \lambda \delta_{ij} \delta_{kl} 
+ 2 \mu \delta_{k(i}\delta_{j)l}
+ \gamma \delta_{ijkl}
\]
where $\lambda=\mu=3A_{hex}^2$, which are the standard Lam\'{e}-parameters of an isotropic elastic medium, $\gamma=0$, which is an elastic coefficient  quantifying any elastic anisotropy, and $\delta_{ijkl}$ is a generalization of the Kronecker-delta symbol which is $1$ if all indices are equal and $0$ otherwise.
These are the same elastic constants as those found in Refs. \cite{elderModelingElasticPlastic2004,skaugenDislocationDynamicsCrystal2018}.
For the 2D hexagonal lattice, $\gamma = 0$, since the tetradic product sum of $\R_{hex}^{(1)}$ is isotropic.
This isotropy, and hence the isotropic elastic properties of the hexagonal lattice, is a direct result of the six-fold rotational symmetry of $\R_{hex}^{(1)}$ \cite{chenMomentIsotropyDiscrete2011}. 

\begin{figure}
    \centering 
    \includegraphics[]{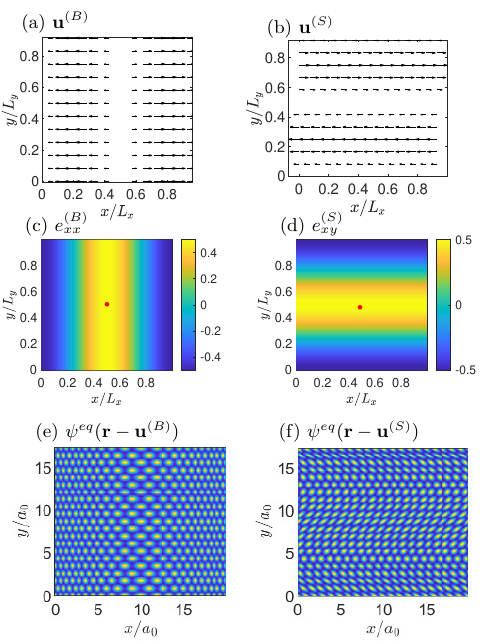}
    \caption{(a) Bulk displacement field $\vec u^{(B)}$ with $e_{xx}^{\max} = 0.5$, (b) Shear displacement field $\vec u^{(S)}$ with $e_{xy}^{\max} = 0.5$,
    (c) $e_{xx}^{(B)}$ and
    (d) $e_{xy}^{(S)} = u_{yx}^{(S)}/2$. 
    The bottom row shows a $20\times 20$ 2D hexagonal PFC distorted by (e) $\vec u^{(B)}$ and (f) $\vec u^{(S)}$. 
    Such highly strained configurations would be prone to melting and nucleation of dislocations if allowed to evolve diffusively, and are included only to illustrate the effects the applied strains.
    The dot in (c-d) marks the central region of the computational domain at which both displacement fields are at their maximal strains. }
    \label{fig:bulk_and_shear_deformation}
\end{figure}

We prepare a $60\times 60$ 2D hexagonal PFC lattice in the one-mode approximation with periodic boundary conditions and lattice vectors reciprocal to $\R_{hex}^{(1)}$, which gives a lattice constant of $a_0=4\pi/\sqrt 3$.
We choose grid spacing $\Delta x=a_0/7$ and $\Delta y =a_0 \sqrt 3/12$ and parameters $r=-0.3$ and $\bar \psi=-0.25$. 
The PFC was deformed by two different displacement fields:
\begin{enumerate}
    \item A bulk displacement field 
    \[\begin{array}{l} 
    u_x^{(B)} = -e_{xx}^{\max}
\frac{L_x}{2 \pi} \sin \left (2\pi \frac{x}{L_x} \right )\\
u_y^{(B)} = 0, \\
\end{array}
\label{eq:bulk_displacement_field}
\]
which corresponds to uniaxial compression/extension in the $(1,0)$-direction.
\item A shear displacement field
\[ 
\begin{array}{l}
u_x^{(S)} = -e_{xy}^{\max} \frac{L_y}{\pi} \sin \left (2\pi \frac{y}{L_y} \right ) \\
u_y^{(S)} = 0, \\
\end{array}
\label{eq:shear_displacement_field}
\]
which corresponds to pure shear in the $(0,1)$-direction. 
\end{enumerate}
Here, $L_x$ and $L_y$ are the lengths of the simulation domain in the $x$- and $y$-direction, respectively.
$e_{xx}^{\max}$ and $e_{xy}^{\max}$ are parameters used to tune the magnitude of the displacement fields.
Figure \ref{fig:bulk_and_shear_deformation} shows the displacement fields and the non-zero components of the strains. 
The displacement fields are constructed in such a way that for $\vec u^{(B)}$ ($\vec u^{(S)}$), the only non-zero component of the strain is $e_{xx}$ ($e_{xy}$), the maximal value of which is $e_{xx}^{\max}$ ($e_{xy}^{\max}$) along the line $x=L_x/2$ ($y=L_y/2$).
For illustrative purposes, we have included the distortion of a $20\times20$ hexagonal lattice by the bulk and shear displacement fields in Figs. \ref{fig:bulk_and_shear_deformation}(e) and (f), respectively.
For the given distortion, the linear elastic energy density is given by 
\[
\Delta f_{el} = \frac{1}{2} C_{ijkl} e_{ij} e_{kl},
\label{eq:linear_elastic_energy}
\]
which for the bulk deformation in the central region, where $e_{xx} = e_{xx}^{\max}, $ gives
\[
\Delta f_{el}^{(B)}
=
 \frac{1}{2}(\lambda+2\mu+\gamma) (e_{xx}^{\max})^2
  .
\label{eq:hex_bulk_elastic_energy}
\]
The free energy density under the bulk deformation, calculated by directly coarse graining Eq. (\ref{eq:free_energy_dens_hex_PFC}), is shown in Fig. \ref{fig:energy_and_stress_of_bulk_and_shear_displacement}(a), where the dashed line indicates the elastic energy of Eq. (\ref{eq:hex_bulk_elastic_energy}).
For the shear deformation, the elastic energy density is given by 
\[
\Delta f_{el}^{(S)} 
=
(2 \mu) (e_{xy}^{\max} )^2 .
\label{eq:hex_shear_elastic_energy}
\]
The free energy under the shear deformation is shown in Fig. \ref{fig:energy_and_stress_of_bulk_and_shear_displacement}(b), where the dashed line indicates the linear elastic energy of Eq. (\ref{eq:hex_shear_elastic_energy}).
We see that the energy stored in the PFC under deformation is correctly accounted for by linear elasticity for small strains.
Figures \ref{fig:energy_and_stress_of_bulk_and_shear_displacement}(c-d) show the values of $h_{xx}^{(2D~hex)}$ and $h_{xy}^{(2D~hex)}$ in the central region (red dot in Figs \ref{fig:bulk_and_shear_deformation}(c-d)) as functions of the applied strains $e_{xx}^{\max}$ and $e_{xy}^{\max}$, respectively, where the dashed line indicates the linear stresses $h_{ij} = C_{ijkl} e_{kl}$. 
We see that the stress in the 2D hexagonal PFC is accounted for by linear elasticity for small strains.


\subsection{2D Square lattice}
\label{sec:2D_square_PFC}

One of the challenges of the PFC formalism is finding suitable free energy functionals that favor a particular lattice symmetry. 
In order to ensure phase stability in the presence of disturbances, such as dislocations or external stresses, the phase-diagram of a proposed free energy must be calculated and parameters chosen so that the desired lattice symmetry minimizes the free energy. 
The postulation of free energy functionals and subsequent calculation of phase-diagrams has thus been the subject of much research \cite{elderModelingElasticPlastic2004,wuPhasefieldCrystalModeling2007,jaatinenExtendedPhaseDiagram2010,wuPhasefieldcrystalModelFcc2010,wuControllingCrystalSymmetries2010,mkhontaExploringComplexWorld2013,emdadiRevisitingPhaseDiagrams2016,elderTwocomponentStructuralPhasefield2018}.
However, a straight-forward generalization of the free energy in Eq. (\ref{eq:free_energy_dens_hex_PFC}) to obtain a square lattice is obtained by noting that the $\L_1$-term was introduced to favor spatial modulations corresponding to the closest modes $\vec q_n^2 = 1$ on the reciprocal lattice. 
For the reciprocal lattice of the square lattice, shown in Fig. (\ref{fig:2D_reciprocal_lattices}), the next-to closest modes $\R_{sq}^{(2)} = \{\vec p_n\}$ with length $\vec p_n^2=2$ will give sizable contributions to the average free energy of Eq. (\ref{eq:free_energy_dens_hex_PFC}). 
Thus, by modifying the free energy to also favor these second modes, one can postulate the free energy functional $F = \int d^2 r \tilde f^{(sq)}$ with
\[
\tilde f^{(sq)} = \frac{1}{2} (\L_1 \L_2 \psi)^2 + \frac{r}{2} \psi^2 + \frac{1}{4} \psi^4,
\label{eq:PFC_square_free_energy}
\]
where 
\[
\L_{X} = X + \nabla^2
\]
is a factor introduced to energetically favor modes of length $\sqrt{X}$.
This free energy was shown to produce a 2D square phase in Ref. \cite{lifshitzTheoreticalModelFaraday1997}. 
The phase diagram for this free energy shows that it has a stable region for the square lattice for a range of values of $r$ \cite{emdadiRevisitingPhaseDiagrams2016}.
Similar to the hexagonal lattice, we get non-zero contributions to $h_{ij}$ from 
\[
\begin{split}
\f_{m_1 m_2} &= 3 \L_1 \L_2 \psi \delta_{m_1 m_2} \\
\f_{m_1 m_2 m_3 m_4} &= \L_1 \L_2 \psi \delta_{(m_1 m_2} \delta_{m_3 m_4)}. 
\end{split} 
\]
Here we are using the general notation for symmetrizing over multiple indices, 
\[
[\cdots]_{(m_1 ... m_\a)} = \frac{1}{\a!}
\sum_{\sigma \in S_\a} [\cdots]_{\sigma(m_1) ... \sigma(m_\a)},
\]
where $S_\a$ is the symmetric group of $\a$ elements. 
This gives
\[\begin{split}
h_{ij}^{(2D~sq)}
&= \avg{M_{ij}^{(2)}} + \avg{M_{ij}^{(4)}}\\
&=
- 6\avg{(\L_1 \L_2 \psi) \d_{ij} \psi} 
- 4\avg{(\L_1 \L_2 \psi) \d_{ijkk} \psi} \\
&=
-2 \avg{(\L_1 \L_2 \psi)(\L_1 + \L_2) \d_{ij} \psi }.
\end{split}\]
Since the square lattice will give sizable contributions also to the second closest modes on the reciprocal lattice, we expand the ground state of the PFC density $\psi$ in the two-mode approximation,
\[
\psi^{eq} = \bar \psi + A_{sq} \sum_{\vec q_n \in \R_{sq}^{(1)}} e^{i\vec q_n \cdot \vec r} + B_{sq} \sum_{\vec p_n \in \R_{sq}^{(2)}} e^{i \vec p_n \cdot \vec r},
\label{eq:PFC_square_two_mode_approximationa}
\]
where $A_{sq}$ and $B_{sq}$ are the equilibrium amplitudes of the modes on the 2D square reciprocal lattice $\R_{sq}^{(1)} = \{ \vec q_{-2}, \vec q_{-1},\vec q_1, \vec q_2, \}$ and $\R_{sq}^{(2)} = \{ \vec p_{-2} \vec p_{-1}, \vec p_1, \vec p_2\}$, respectively, where
\[
\begin{array}{ll}
    \vec q_1 = (1,0)   & \vec p_1 = (1,-1) \\
    \vec q_2 = (0,1)   & \vec p_2 = (1,1),
\end{array}
\label{eq:Rsq}
\]
$\vec q_{-n} = -\vec q_{n}$ and $\vec p_{-n} = -\vec p_{n}$, see Fig. \ref{fig:2D_reciprocal_lattices}(b).
$A_{sq},B_{sq}$ are determined by minimization of the free energy.
By deforming the two-mode approximation by a displacement field $\vec u$, we find (Appendix  \ref{sec:appendix_square_mode_expansion})
\[\begin{split}
h_{ij}^{(2D~sq)} &= 4 A_{sq}^2 \d_k u_l   \sum_{\vec q_n \in \R_{sq}^{(1)}} q_{ni} q_{nj} q_{nl} q_{nk} \\
&+ 4 B_{sq}^2 \d_k u_l  \sum_{\vec p_n \in \R_{sq}^{(2)}} p_{ni} p_{nj} p_{nk} p_{nl}.
\end{split}\]
The tetradic product sums are given by
\[
\sum_{\vec q_n \in \R_{sq}^{(1)}} q_{ni} q_{nj} q_{nl} q_{nk}
=
2 \delta_{ijkl},
\]
\[
\sum_{\vec p_n \in \R_{sq}^{(2)}} p_{ni} p_{nj} p_{nk} p_{nl}
=
4 (\delta_{ij}\delta_{kl} + 2 \delta_{k(i} \delta_{j)l} - 2\delta_{ijkl}).
\]
This gives elastic constants $\lambda = \mu = 16B_{sq}^2$ and $\gamma = 8 A_{sq}^2- 32 B_{sq}^2$,
which match those found in previous work for the 2D square PFC lattice \cite{wuPhasefieldcrystalModelFcc2010}. 
Ref. \cite{chenMomentIsotropyDiscrete2011} showed that for a collection of vectors that have a four-fold symmetry, such as $\R_{sq}^{(1)}$, only rank 2 moment tensors are identically isotropic, which explains the anisotropic nature of the 2D square PFC model. 
However, the model exhibits isotropic elasticity ($\gamma=0$) in the case of $B_{sq}=A_{sq}/2$, which can be solved with the equilibrium conditions on the amplitudes to give
\[
r=-\frac{25}{3}\P^2,
\]
which falls within the region of a stable square lattice phase, indicating that a stable configuration for the isotropic square crystal does exist \cite{emdadiRevisitingPhaseDiagrams2016}. 

We prepare a $60\times 60$ 2D square PFC lattice in the two-mode approximation on periodic boundaries with lattice vectors reciprocal to $\R_{sq}^{(1)}$, which gives a lattice constant of $2\pi$. We choose grid spacings $\Delta x=\Delta y = a_0/7$ and parameters $r=-0.3$ with $\bar \psi=-0.25$.
The PFC is deformed according to the displacement fields of Eqs. (\ref{eq:bulk_displacement_field}-\ref{eq:shear_displacement_field}), for which the elastic energy density again scales with the square of the strain as in Eqs. (\ref{eq:hex_bulk_elastic_energy}-\ref{eq:hex_shear_elastic_energy}), and is shown in Fig. (\ref{fig:energy_and_stress_of_bulk_and_shear_displacement}).


\subsection{3D bcc lattice}
\begin{figure*}
    \centering
    \includegraphics[]{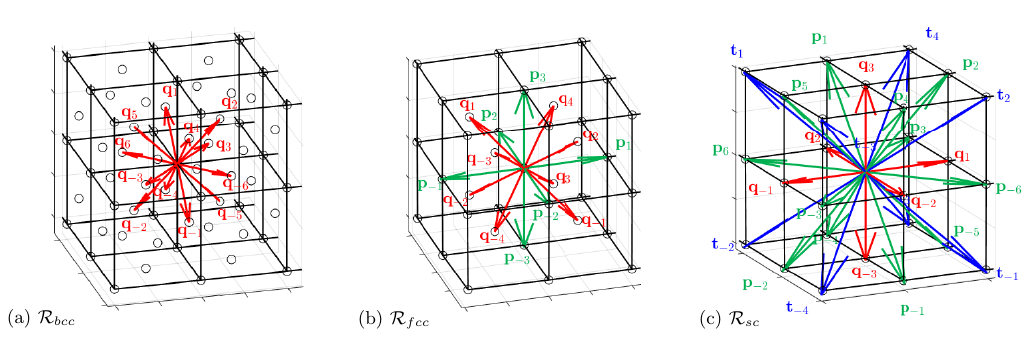}
    \caption{Reciprocal lattices of the 3D crystal structures. (a) Modes $\R_{bcc}$ of the bcc reciprocal lattice: $\R_{bcc}^{(1)} = \{\vec q_{n} \}$ are the closest reciprocal lattice modes ($\vec q_n^2 = 1$).  (b) Modes $\R_{fcc}$ of the fcc reciprocal lattice: $\R_{fcc}^{(1)} = \{\vec q_{n} \}$ are the closest reciprocal lattice modes ($\vec q_n^2 = 1$) and $\R_{fcc}^{(4/3)} = \{\vec p_n\}$ are the next-to-closest modes ($\vec p_n^2 = 4/3$). 
    (c) Modes $\R_{sc}$ of the sc reciprocal lattice: $\R_{sc}^{(1)} = \{\vec q_{n} \}$ are the closest reciprocal lattice modes ($\vec q_n^2 = 1$), $\R_{sc}^{(2)} = \{\vec p_n\}$ are the next-to-closest modes ($\vec p_n^2 = 2$), and
    $\R_{sc}^{(3)} = \{\vec t_n\}$ are the third closest modes ($\vec t_n^2 = 3$).}
    \label{fig:3D_reciprocal_lattices}
\end{figure*}

In three dimensions, for a suitable range of parameters, the equilibrium configuration of the original PFC model with the free energy given in Eq. (\ref{eq:free_energy_dens_hex_PFC}) is that  of a bcc lattice \cite{wuPhasefieldCrystalModeling2007}. 
The associated free energy  is $F = \int d^3 r \tilde f^{(bcc)}$ with 
\[
\tilde f^{(bcc)} = \frac{1}{2} (\L_1\psi)^2 + \frac{r}{2} \psi^2 +\frac{1}{4} \psi^4.
\label{eq:PFC_bcc_free_energy}
\]
The non-zero free energy derivatives are the same as for the 2D hexagonal lattice, so
\[
h_{ij}^{(3D~bcc)}= -2 \avg{(\L_1 \psi) \d_{ij} \psi}.
\]
The elastic constants are calculated by expanding the ground state in the one-mode expansion 
\[
\psi^{eq} = \P + A_{bcc} \sum_{\vec q_n \in \R_{bcc}^{(1)}} e^{i\vec q_n \cdot \vec r}, 
\]
where $A_{bcc}$ is the equilibrium amplitude of the closest modes on the reciprocal lattice for the bcc lattice. 
The latter can be chosen as $\R_{bcc}^{(1)} = \{\vec q_{-6},...,\vec q_{-1},\vec q_1,...,\vec q_6\}$, where 
\[
\begin{array}{ll}
\vec q_1 = (0,1,1)/\sqrt 2 &  \vec q_4 = (0,-1,1)/\sqrt 2\\
\vec q_2 = (1,0,1)/\sqrt 2 & \vec q_5 = (-1,0,1)/\sqrt 2 \\
\vec q_3 = (1,1,0)/\sqrt 2 & \vec q_6 = (-1,1,0)/\sqrt 2\\
\end{array}
\]
and $\vec q_{-n} = -\vec q_n$, see Fig. \ref{fig:3D_reciprocal_lattices}(a). 
$A_{bcc}$ is found by minimization of the free energy.
By straining the equilibrium state in the one-mode approximation, we find (Appendix \ref{sec:appendix_bcc_mode_expansion})
\[
h^{(3D~bcc)}_{ij} = 4 A_{bcc}^2 \d_k u_l \sum_{\vec q_n \in \R_{bcc}^{(1)}} q_{ni} q_{nj} q_{nk} q_{nl}.
\]
Now, using the tetradic product sum of $\R_{bcc}^{(1)}$,
\[
\sum_{\vec q_n \in \R_{bcc}^{(1)}} q_{ni} q_{nj} q_{nk} q_{nl} = (\delta_{ij} \delta_{kl}  + 2 \delta_{k(i} \delta_{j) l} - \delta_{ijkl}),
\]
we find the elastic constants $\lambda = \mu = 4A_{bcc}^2$ and $\gamma = -4 A_{bcc}^2$, 
as found for the fcc PFC in Ref. \cite{wuPhasefieldcrystalModelFcc2010}.
Since $\gamma\neq 0$, this PFC model will not exhibit isotropic elasticity.

We prepare a $60\times 60\times 5$ bcc PFC lattice in the one-mode approximation on periodic boundaries with lattice vectors reciprocal to $\R_{bcc}^{(1)}$, which gives a lattice constant of $2\pi\sqrt 2$. 
We choose a grid spacing of $\Delta x=\Delta y = \Delta z=a_0/7$ and parameters $r=-0.3$ with $\bar \psi=-0.325$.
The PFC is deformed according to the displacement fields of Eqs. (\ref{eq:bulk_displacement_field}-\ref{eq:shear_displacement_field}), extended to three dimensions with $u_z^{(B)} = u_z^{(S)} = 0$, for which the elastic energy density scales with the square of the strain as in Eqs. (\ref{eq:hex_bulk_elastic_energy}-\ref{eq:hex_shear_elastic_energy}). 
The results are shown in Fig. (\ref{fig:energy_and_stress_of_bulk_and_shear_displacement}).


\subsection{3D fcc lattice}
The introduction of $\L_2$ in the free energy functional for the 2D square PFC was motivated by contributions from the next-to closest reciprocal modes. 
Similarly, the inclusion of a differential operator that favors density waves at reciprocal modes of length $\sqrt {4/3}$ might produce a stable fcc lattice. 
This motivates the following form of the PFC fcc model: $F = \int d^3 r \tilde f^{(fcc)}$ with 
\[
\tilde f^{(fcc)} = \frac{1}{2} (\L_1 \L_{\frac{4}{3}} \psi)^2 + \frac{r}{2} \psi^2 +\frac{1}{4} \psi^4,
\label{eq:PFC_fcc_free_energy}
\]
which has been shown to produce a stable fcc phase \cite{wuPhasefieldcrystalModelFcc2010}. 
The non-zero derivatives of the free energy density are
\[
\begin{split}
\f_{m_1 m_2} &= \frac{7}{3} \L_1 \L_{\frac 4 3} \psi \delta_{m_1 m_2} \\
\f_{m_1 m_2 m_3 m_4} &= \L_1 \L_{\frac 4 3} \psi \delta_{(m_1 m_2} \delta_{m_3 m_4)},
\end{split} 
\]
so
\[\begin{split}
h_{ij}^{(3D~fcc)} &= \avg{M_{ij}^{(2)}} + \avg{M_{ij}^{(4)}} \\
&= -2 \avg{(\L_1 \L_{\frac 4 3} \psi ) (\L_1 + \L_{\frac 4 3}) \d_{ij} \psi  }.
\end{split}
\]
The elastic constants are calculated by expanding the ground state in the two-mode expansion 
\[
\psi^{eq} = \P + A_{fcc} \sum_{\vec q_n \in \R_{fcc}^{(1)}} e^{i\vec q_n \cdot \vec r} + B_{fcc} \sum_{\vec p_n \in \R_{fcc}^{(4/3)}} e^{i\vec p_n \cdot \vec r},
\]
where $A_{fcc}$ and $B_{fcc}$ are the equilibrium amplitudes of the modes on the reciprocal lattice lattice $\R_{fcc}^{(1)} = \{\vec q_{-4},...,\vec q_{-1},\vec q_1,...,\vec q_4\}$ and $\R_{fcc}^{(4/3)} = \{\vec p_{-3},...,\vec p_{-1},\vec p_1,...,\vec p_3\}$, respectively, where 
\[
\begin{array}{ll}
    \vec q_1 = (-1,1,1)/\sqrt 3  & \vec p_1 = (2,0,0)/\sqrt 3  \\
    \vec q_2 = (1,-1,1)/\sqrt 3 & \vec p_2 = (0,2,0)/\sqrt 3\\
    \vec q_3 = (1,1,-1)/\sqrt 3 & \vec p_3 = (0,0,2)/\sqrt 3\\
    \vec q_4 = (1,1,1)/\sqrt 3 &,
\end{array}
\]
with $\vec q_{-n} = -\vec q_n$ and $\vec p_{-n} = -\vec p_n$, see Fig. \ref{fig:3D_reciprocal_lattices}(b). 
$A_{fcc}$ and $B_{fcc}$ are found by minimization of the free energy. 
By deforming the two-mode approximation by a displacement field $\vec u$, we find (Appendix \ref{sec:appendix_fcc_mode_expansion})
\[\begin{split}
h_{ij}^{(3D~fcc)}
&=
\frac{4}{9} A_{fcc}^2 \d_k u_l \sum_{\vec q_n \in \R_{fcc}^{(1)}}
q_{nk} q_{nl} q_{ni} q_{nj} \\
&+
\frac{4}{9} B_{fcc}^2 \d_k u_l \sum_{\vec p_n \in \R_{fcc}^{(4/3)}}
p_{nk} p_{nl} p_{ni} p_{nj}.
\end{split}\]
The tetradic product sums are given by
\[
\sum_{\vec q_n \in \R_{fcc}^{(1)}}q_{ni} q_{nj} q_{nk} q_{nl} 
=
\frac{8}{9} (\delta_{ij} \delta_{kl}  + 2 \delta_{k(i} \delta_{j) l} - 2\delta_{ijkl}),
\]
\[
\sum_{\vec p_n \in \R_{fcc}^{(4/3)}} p_{ni} p_{nj} p_{nk} p_{nl} = \frac{32}{9} \delta_{ijkl},
\]
which gives elastic constants $\lambda = \mu = \frac{32}{81} A_{fcc}^2$ and $\gamma = \frac{64}{81}(2B_{fcc}^2 - A_{fcc}^2)$,
as found for the fcc PFC in Ref. \cite{wuPhasefieldcrystalModelFcc2010}.
The expression for $\gamma$ shows that the fcc lattice exhibits isotropic elasticity ($\gamma=0$) if $B_{fcc}=A_{fcc}/\sqrt 2$, which solved with the equilibrium condition on the amplitudes gives
\[
r = -\frac{255 \P^2}{49},
\]
which falls within the region of a stable fcc phase, indicating that a stable configuration of isotropic elasticity for the fcc lattice does exist \cite{wuPhasefieldcrystalModelFcc2010}.

We prepare a $60\times60\times 5$ fcc PFC lattice in the two-mode approximation on periodic boundaries with lattice vectors reciprocal to  $R_{fcc}^{(1)}$, which gives a lattice constant of $2\pi \sqrt 3$. We choose a grid spacing of $\Delta x = \Delta y = \Delta z = a_0/11$ and parameters $r=-0.3$ with $\bar \psi=-0.3$. 
We perform the same distortion of the fcc PFC by the bulk and shear displacement fields, for which the elastic energy density scales with the square of the strain as in Eqs. (\ref{eq:hex_bulk_elastic_energy}-\ref{eq:hex_shear_elastic_energy}). 
The results are shown in Fig. \ref{fig:energy_and_stress_of_bulk_and_shear_displacement}. 


\subsection{3D Sc lattice}

Extending the idea of favoring modes of the reciprocal lattice to achieve other symmetries, a natural generalization of the free energy in Eq. (\ref{eq:PFC_fcc_free_energy}) can be chosen as follows. 
The three sets of modes that are closest to the origin on the sc reciprocal lattice are given by $\R_{sc}^{(1)} = \{\vec q_{-3},...,\vec q_{-1},\vec q_1,...,\vec q_3\}$, $\R_{sc}^{(2)} = \{\vec p_{-6},...,\vec p_{-1},\vec p_1,...,\vec p_6\}$ and $\R_{sc}^{(3)} = \{\vec t_{-4},...,\vec t_{-1},\vec t_1,...,\vec t_4\}$, where   
\[
\begin{array}{lll}
    \vec q_1 = (1,0,0)  & \vec p_1 = (0,1,1) & \vec t_1 = (-1,1,1)  \\
    \vec q_2 = (0,1,0) & \vec p_2 = (1,0,1) & \vec t_2 = (1,-1,1)\\
    \vec q_3 = (0,0,1) & \vec p_3 = (1,1,0) & \vec t_3 = (1,1,-1)\\
    & \vec p_4 = (0,-1,1) & \vec t_4 = (1,1,1)\\
    & \vec p_5 = (-1,0,1) & \\
    & \vec p_6 = (-1,1,0) &, \\
\end{array}
\]
with $\vec q_{-n} = -\vec q_n$, $\vec p_{-n} = -\vec p_n$ and  $\vec t_{-n} = -\vec t_n$, see Fig. \ref{fig:3D_reciprocal_lattices}(c). 
Thus, a way to explicitly favor the simple cubic structure is to introduce the free energy $F=\int d^3 r \tilde f^{(sc)}$ where
\[
\tilde f^{(sc)}
=
\frac{1}{2} (\L_1 \L_2 \L_3 \psi)^2 +\frac{r}{2} \psi^2 +\frac{1}{4} \psi^4,
\label{eq:SC_free_energy_density}
\]
 which corresponds to a special case of the multimode PFC expansion of Ref. \cite{mkhontaExploringComplexWorld2013}.
In order to ensure a stable sc phase, one needs to consider the competing symmetries, calculate the average free energy density for each symmetry and find coexistence regions by Maxwell construction. 
We leave this task for future work, since the interest of the current paper is to find a point in configuration space $(r,\P)$ for which the sc phase is stable.
This can be done by searching for parameters $(r,\P)$ for which a random initial condition  condenses into the simple cubic phase.
We have found $(r,\P)=(-0.3,-0.325)$ to be such a point.
Expanding the ground state in the three-mode expansion,
\[\begin{split}
\psi^{eq} = \P + A_{sc} \sum_{\vec q_n \in \R_{sc}^{(1)}} e^{i\vec q_n \cdot \vec r} + B_{sc} \sum_{\vec p_n \in \R_{sc}^{(2)}} e^{i\vec p_n \cdot \vec r} \\
+C_{sc} \sum_{\vec t_n \in \R_{sc}^{(3)}} e^{i\vec t_n \cdot \vec r},
\end{split}\]
inserting into the free energy Eq. (\ref{eq:SC_free_energy_density}), and averaging over a unit cell of size $(2\pi)^3$, gives 
\[\begin{split}
\avg{\tilde f^{(sc)}}
&=
\frac{1}{2} (6 {\bar \psi})^2 + \frac{r}{2}{\bar \psi}^2+ \frac{1}{4} {\bar \psi}^4
\\
&+\frac{45}{2}A_{sc}^4 + 288 A_{sc}^2 B_{sc}^2 + 135 B_{sc}^4 + 48 A_{sc}^3 C_{sc} 
\\
&+ 432 A_{sc} B_{sc}^2 C_{sc} + 108 A_{sc}^2 C_{sc}^2 + 324 B_{sc}^2 C_{sc}^2 
\\
&+ 54 C_{sc}^4 + 72 A_{sc}^2 B_{sc} {\bar \psi} + 48 B_{sc}^3 {\bar \psi} 
\\
&+ 144 A_{sc}B_{sc} C_{sc} {\bar \psi}  + 9 A_{sc}^2 {\bar \psi}^2 + 18 B_{sc}^2 {\bar \psi}^2 
\\
&+ 12 C_{sc}^2 {\bar \psi}^2  + 3 A_{sc}^2 r + 6 B_{sc}^2 r + 4 C_{sc}^2 r, 
\end{split}\]
where the equilibrium values of $A_{sc}, B_{sc}, C_{sc}$ are determined by minimization of the free energy. 
From Eq. (\ref{eq:SC_free_energy_density}), we get 
\[
\begin{array}{ll}
\f_{m_1 m_2} &= 11 \L_1 \L_2 \L_3 \psi \delta_{m_1 m_2}\\
\f_{m_1 m_2 m_3 m_4} &= 6 \L_1 \L_2 \L_3 \psi \delta_{(m_1 m_2} \delta_{m_3 m_4)}\\
\f_{m_1m_2m_3m_4m_5m_6} &= \L_1 \L_2 \L_3 \psi \delta_{(m_1 m_2} \delta_{m_3 m_4} \delta_{m_5 m_6)},
\end{array}
\]
which gives
\[\begin{split}
h_{ij}^{(3D~sc)}
&=
\avg{M_{ij}^{(2)}} 
+\avg{M_{ij}^{(4)}} 
+\avg{M_{ij}^{(6)}} \\
&=
-2\avg{(\L_1 \L_2 \L_3 \psi)(\L_2 \L_3 + \L_1 \L_3 + \L_1 \L_2) \d_{ij} \psi}.
\end{split}\]
By deforming the three-mode expansion by a displacement field $\vec u$, we find (Appendix  \ref{sec:appendix_sc_mode_expansion})
\[\begin{split}
h_{ij}^{(3D~sc)}
=
16 A_{sc}^2 \d_k u_l \sum_{\vec q_n \in \R_{sc}^{(1)}} q_{ni} q_{nj} q_{nk} q_{nl} \\
+ 4 B_{sc}^2 \d_k u_l \sum_{\vec p_n \in \R_{sc}^{(2)}} p_{ni} p_{nj} p_{nk} p_{nl} \\
+ 16 C_{sc} \d_{k} u_l \sum_{\vec t_n \in \R_{sc}^{(3)}} t_{ni} t_{nj} t_{nk} t_{nl},
\end{split}\]
which after using the tetradic product sums 
\[
\sum_{\vec q_n \in \R_{sc}^{(1)}}
q_{ni} q_{nj} q_{nk} q_{nl}
= 2 \delta_{ijkl}
\]
\[
\sum_{\vec p_n \in \R_{sc}^{(2)}}
p_{ni} p_{nj} p_{nk} p_{nl}
=
4(\delta_{ij} \delta_{kl}
+
 2 \delta_{k(i} \delta_{j)l}
- \delta_{ijkl})
\]
\[
\sum_{\vec t_n \in \R_{sc}^{(3)}}
t_{ni} t_{nj} t_{nk} t_{nl}
=
8 (\delta_{ij} \delta_{kl}
+
  2 \delta_{k(i} \delta_{j)l}
-
2 \delta_{ijkl})
\]
gives elastic constants $\lambda = \mu = 16 B_{sc} + 128 C_{sc}^2$ and $\gamma = 32_{sc} A^2 - 16 B_{sc}^2 - 256 C_{sc}^2$.
The sc PFC would exhibit isotropic elasticity for $\gamma=0$.
An exact expression for the domain of isotropic elasticity could in principle be obtained, as for the previous symmetries, by solving the equilibrium condition on the amplitudes with the additional constraint of $\gamma = 0$. 
This is a set of $4$ quartic equations with $5$ unknowns ($r$,$\psi$,$A_{sc}$,$B_{sc}$,$C_{sc}$) which must be solved simultaneously in order to express the regime of isotropic elasticity.
Using computational software  \cite{Mathematica} suggests that no closed-form solution exists as in the case of the lattices in which only two amplitudes were needed.
However, by numerically solving the equations, we have found the following relation for an isotropic domain
\[
r \approx -2.10144 \P^2.
\]
A PFC model of simple cubic ordering was also considered in Ref. \cite{greenwoodFreeEnergyFunctionals2010} by inclusion of exponential peaks in the correlation function derived from the free energy and in Ref. \cite{wangAngleadjustableDensityField2018} by adding higher order gradients in the free energy to account for an orientation dependent interaction. 
However, to our knowledge, the elastic constants for the particular free energy density of Eq. (\ref{eq:SC_free_energy_density}), have not previously been derived. 
The PFC model employed in Ref. \cite{wangAngleadjustableDensityField2018} has the free energy given in terms of the density field and its derivatives. 

We prepare a $60\times60\times 5$ sc PFC lattice in the three-mode approximation on periodic boundaries with lattice vectors reciprocal to  $R_{sc}^{(1)}$, which gives a lattice constant of $2\pi$. We choose a grid spacing of $\Delta x = \Delta y = \Delta z = a_0/7$ and parameters $r=-0.3$ with $\bar \psi=-0.325$. 
We perform the same distortion of the sc PFC by the bulk and shear displacement fields, for which the elastic energy density scales with the square of the strain as in Eqs. (\ref{eq:hex_bulk_elastic_energy}-\ref{eq:hex_shear_elastic_energy}). 
The results are shown in Fig. \ref{fig:energy_and_stress_of_bulk_and_shear_displacement}. 

\begin{figure*}
    \centering
    \includegraphics[]{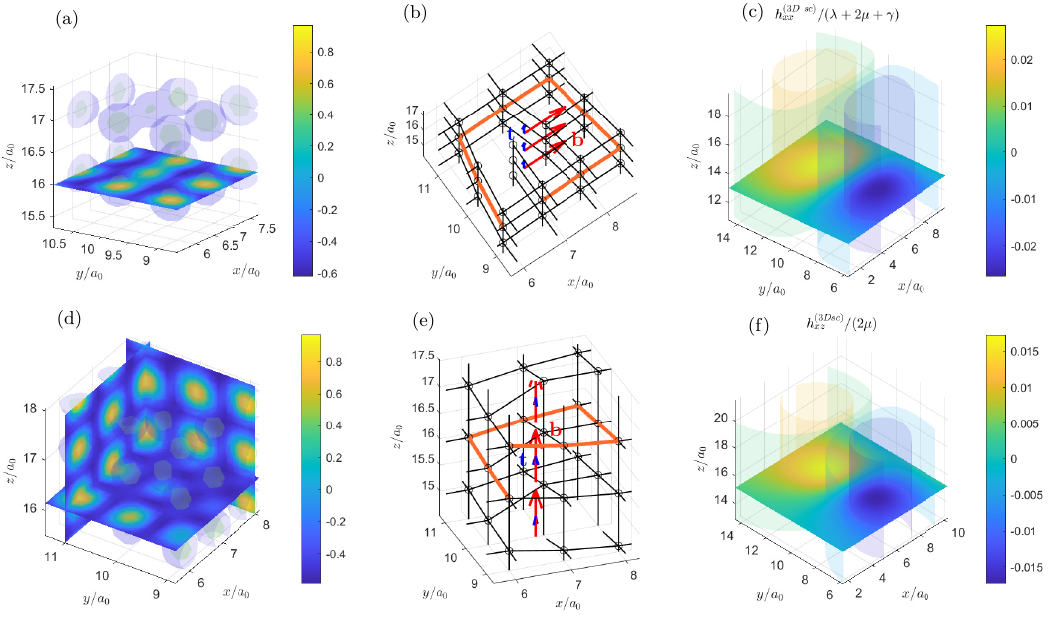}
    \caption{PFC density field $\psi$ for (a) an edge dislocation with Burgers vector $\vec b=a_0(1,0,0)$ and tangent vector $\vec t=(0,0,1)$ and (d) a screw dislocation with Burgers vector $\vec b=a_0(0,0,1)$ and tangent vector $\vec t=(0,0,1)$.
    The $\psi$-field is shown as transparent iso-surfaces of the density $\psi$ with extra inserted 2D density plots for selected planes. (b) and (e) show the peaks of (a) and (d), respectively, connected by lines for illustration. 
    The orange line shows the Burgers circuit with the corresponding closure failure.
    (c) and (f) show the largest components of the stress fields for each dislocation.
    }
    \label{fig:SC_dislocations_illiustration}
\end{figure*}

To exemplify that the formalism extends to defected lattices, we include in Fig. (\ref{fig:SC_dislocations_illiustration}) a 3D sc PFC structure in the presence of an edge (screw) dislocation in panel (a) (panel (c)), and its associated stress field in panel (c) (panel (f)). 
The PFC configuration was prepared by multiplying the complex amplitudes of the PFC by dislocation charges followed by a brief period of relaxation, as explained in Ref. \cite{skaugenDislocationDynamicsCrystal2018}.
Even though dislocations are singular objects (they are phase singularities for the complex amplitudes), the density field $\psi$ and its corresponding macroscopic stress field remain well-defined and smooth, without any core regularization method.
It is interesting that the largest value of the stresses $\sim 0.02$ (in dimensionless units) at the dislocation cores is still nominally on the linear stress-strain curves (see Fig. \ref{fig:energy_and_stress_of_bulk_and_shear_displacement}), even though the physics at the core is fundamentally different. 
This suggests that non-linear elastic effects may not be necessary to capture the near-core deformations, at least in the PFC models, and that the deviations from linear elasticity are due directly to the lattice incompatibility. 
However at present, the stress profile obtained around dislocations are not constrained to mechanical equilibrium, i.e. $\partial_i h_{ij}\neq0$, and thus they are not readily comparable with analytical stress profiles from continuum mechanics. 
The problem of extending the method proposed in Ref. \cite{skaugenSeparationElasticPlastic2018} to anisotropic 2D and 3D PFC lattices to constrain the diffusive dynamics of the PFC models to mechanical equilibrium is an open area of research.  

\section{Conclusion}
\label{sec:conclusion}
To summarize, we have presented a versatile method of computing the macroscopic stress tensor for ordered systems starting from a microscopic field description, where there is an intrinsic length scale (hence a finite intrinsic representative volume element) related to lattice periodicity.
Within a generic field theory of Ginzburg-Landau type, where the system is described by a free energy that depends on a one-body density field (or order parameter) and an arbitrary number of its gradients, we have derived a general formula for the stress tensor given by Eq. (\ref{eq:general_stress_tensor}). 
Upon coarse graining to continuum scales, we obtain the macroscopic stress tensor in the linear regime which describes the state of deformation of the ordered phase as a continuum elastic medium. 

By adopting the PFC formalism to describe crystals, we have derived the stress tensor for different lattice symmetries in two and three dimensions. 
In particular, we focused on the hexagonal and square lattices in two dimensions and bcc, fcc and sc lattices in three dimensions. 
For simplicity, we only looked at the equilibrium defect-free crystal configurations to derive the elastic constants at constant macroscopic density.
We show how the crystal symmetries constrain the tetradic product sums which determine the number of elastic constants and their values. 
For instance, the isotropic elasticity of the 2D hexagonal PFC model is due to the six-fold symmetry of its reciprocal lattice.
For the other PFC models, we have found regions in parameter space where the elastic behavior is expected to be isotropic, except for the bcc PFC model, which is always anisotropic. 
Using numerical simulations, we verified the predicted linear elastic response of all models and found that the 3D fcc lattice model quickly enters a nonlinear elastic regime upon compression/extension.

The formalism developed in this paper can be extended to non-equilibrium fields, with the particular example of a hexagonal lattice in 2D already discussed in Ref. \cite{skaugenSeparationElasticPlastic2018}. 
In this case, one is concerned with the evolution of defected ordered systems whose defects move on timescales much longer than the fast relaxation to elastic equilibrium. 
Our stress tensor formula can be readily applied also in the presence of defects, and can in fact be used to indicate the nucleation and position of defects as discussed in Ref. \cite{skogvollDislocationNucleationPhasefield2021}.
While the analysis of ordered systems is limited to linear elasticity, we have shown that due to the emergent regularization of the dislocation core in the PFC model, non-linear strain effects may not be necessary to capture the near-core deformations.
The method also is applicable to other Ginzburg-Landau theories with an emergent length scale such as mean-field descriptions of liquid crystals and active matter.

\begin{acknowledgments}
We thank Jorge Vi\~nals for many stimulating discussions and his valuable comments at various stages during this work.
We also thank 
Marco Salvalaglio for his valuable input on the manuscript. 
\end{acknowledgments}


\appendix
\begin{widetext}
\section{Proof of the force balance equation}
\label{sec:appendix_proof_of_generalization}

The microscopic stress tensor is given by 
\[
\tilde \sigma_{ij} = (\tilde f-\tilde \mu_c \n)\delta_{ij} + \tilde h_{ij},
\]
where 
\[
\tilde \mu_c = 
\frac{\delta F}{\delta \n}
=
\sum_{\a=0}^\infty (-1)^\a \d_{m_1...m_\a} \f_{m_1...m_\a}
\label{eq:mu_c_in_terms_of_f_hat}.
\]
To show that $\d_i \tilde \sigma_{ij} = -\n \d_j \tilde \mu_c$, we need to show that $\d_i \tilde h_{ij} =  \tilde \mu_c \d_j \n - \d_j \tilde f$.
\[\begin{split}
    \d_i \tilde h_{ij} 
    &=  \d_i \left (\sum_{\a=1}^\infty
     \sum_{\b=1}^{\a} (-1)^\b )(\d_{m_1 ... m_{\beta-1}} \f_{m_1...m_{\a-1}i})\d_{j m_\b ... m_{\a-1}} \n \right )\\
     &=
   \sum_{\a=1}^\infty
     \sum_{\b=1}^{\a} 
     (-1)^\b (\d_{im_1 ... m_{\beta-1}} \f_{m_1...m_{\a-1}i}) 
     \d_{j m_\b ... m_{\a-1}} \n   
     +
     \sum_{\a=1}^\infty
     \sum_{\b=1}^{\a} 
     (-1)^\b (\d_{m_1 ... m_{\beta-1}} \f_{m_1...m_{\a-1}i})
     \d_{ij m_\b ... m_{\a-1}} \n   \\
      &=
      \sum_{\a=1}^\infty
     (-1)^\a (\d_{m_1 ... m_{\a-1}i} \f_{m_1...m_{\a-1}i})
     \d_j \n
     +
     \underset{(1)}{\underbrace{\sum_{\a=1}^\infty
     \sum_{\b=1}^{\a-1} 
     (-1)^\b (\d_{m_1 ... m_{\beta-1}i} \f_{m_1...m_{\a-1}i}) 
     \d_{j m_\b ... m_{\a-1}} \n }} \\
     &-\sum_{\a=1}^\infty
      \f_{m_1...m_{\a-1}i}
     \d_{ m_1 ... m_{\a-1}i j} \n 
     +
     \underset{(2)}{\underbrace{
     \sum_{\a=1}^\infty
     \sum_{\b=2}^{\a} 
     (-1)^\b (\d_{m_1 ... m_{\beta-1}} \f_{m_1...m_{\a-1}i})
     \d_{j i m_\b ... m_{\a-1}} \n  }}.
     \label{eq:proof_eq1}
\end{split}\]
In the last equality, the second term on the first line (1) cancels the second term on the second line (2) as is seen by starting with (1), switching dummy indices $m_\b\leftrightarrow i$, using that $\f_{m_1...m_\a}$ is symmetric under the interchange of indices, and adjusting summation limits as follows
\begin{multline}
\sum_{\a=1}^\infty
     \sum_{\b=1}^{\a-1} 
     (-1)^\b (\d_{m_1 ... m_{\beta-1}i} \f_{m_1...m_{\a-1}i}) 
     \d_{j m_\b ... m_{\a-1}} \n \\
     =
     \sum_{\a=1}^\infty
     \sum_{\b=1}^{\a-1} 
     (-1)^\b (\d_{m_1 ... m_{\beta-1}m_\beta} 
     \f_{m_1...m_{\b-1} i m_{\b+1}...m_{\a-1} m_{\b}})
     \d_{ji m_{\b+1} ... m_{\a-1}} \n \\
     =
     -
     \sum_{\a=1}^\infty
     \sum_{\b=2}^{\a} 
     (-1)^\b (\d_{m_1 ... m_{\beta-1}} \f_{m_1...m_{\a-1}i})
     \d_{ji m_{\b} ... m_{\a-1}} \n \\
\end{multline}
Thus, adding and subtracting $\f \d_j \n $ from Eq. (\ref{eq:proof_eq1}), and renaming the dummy index $i\rightarrow m_\a$, we get
\[\begin{split}
    \d_i \tilde h_{ij} 
    &=
    \f \d_j \n 
    +
    \sum_{\a=1}^\infty
     (-1)^\a (\d_{m_1 ... m_{\a}} \f_{m_1...m_{\a}})
     \d_j \n
     -   \f \d_j \n 
      -\sum_{\a=1}^\infty
      \f_{m_1...m_{\a}}
     \d_{ m_1 ... m_{\a} j} \n  \\
     &=\sum_{\a=0}^\infty
     (-1)^\a (\d_{m_1 ... m_{\a}} \f_{m_1...m_{\a}})
     \d_j \n
     - 
     \left (
     \f \d_j \n
      +\sum_{\a=1}^\infty
      \f_{m_1...m_{\a}}
     \d_{ m_1 ... m_{\a} j} \n 
     \right )
     =
     \tilde \mu_c \d_j \n - \d_j \tilde f,
\end{split}\]
where we have used that by the chain rule, we have
\[
\d_j \tilde f
=
\f \d_j \n + \f_{m_1} \d_{m_1 j} \n + ...
=
\f \d_j \n
+
\sum_{\a=1}^\infty
\f_{m_1 ... m_\a} \d_{m_1 ... m_\a j} \n.
\]


\subsection{The combinatorial factor}
\label{sec:appendix_combinatorial_factor}

When defining $\f_{m_1,...,m_\a}$, the combinatorial factor $N(\{m_1,\ldots m_\a\})$ appears due to arbitrary gradients of $\n$ not being independent, for example  $\d_{xy} \n = \d_{yx} \n$. 
Therefore, in the Taylor expansion to first order, we should include only one of each term. 
To illustrate, assume that the free energy $F[\n, \{ \d_i \n\}, \{\d_{ij} \n\} ]$ is given in terms of $\n$, which is a field in 2 dimensions, and up to its second order gradients. 
In this case, the variation of $F$ is given by 
\[
\begin{split}
\delta F 
&= \int d^2 r\left (
\frac{\d f}{\d \n} \delta \n 
+\frac{\d f}{\d (\d_{x} \n)} \delta (\d_{x} \n) 
+\frac{\d f}{\d (\d_{y} \n)} \delta (\d_{y} \n) 
+\frac{\d f}{\d (\d_{xx} \n)} \delta (\d_{xx} \n )
+\frac{\d f}{\d (\d_{yy} \n)} \delta (\d_{yy} \n )
+\frac{\d f}{\d (\d_{xy} \n)} \delta (\d_{xy} \n)   
\right ) \\
&=\int d^2 r\left (
\frac{\d f}{\d \n} \delta \n 
+\frac{\d f}{\d (\d_{x} \n)} \delta (\d_{x} \n )
+\frac{\d f}{\d (\d_{y} \n)} \delta (\d_{y} \n )
+\frac{\d f}{\d (\d_{xx} \n)} \delta (\d_{xx} \n )
+\frac{\d f}{\d (\d_{yy} \n)} \delta (\d_{yy} \n )
\right . \\ 
&
\left .
\hspace{9.9cm} +\frac{1}{2} \frac{\d f}{\d (\d_{xy} \n)} \delta (\d_{xy} \n )
+\frac{1}{2} \frac{\d f}{\d (\d_{yx} \n)} \delta (\d_{yx} \n)
 \right ) \\
&=
\int d^2 r \left (\f \delta \n + \f_{m_1} (\delta \d_{m_1} \n) + \f_{m_1 m_2} \delta (\d_{m_1 m_2} \n) \right ),
\end{split}
\]
Where the combinatorial factor of $N(\{x,y\}) = 1/2$ was needed to write the sum over all indices. 
This gives after integration by parts (ignoring boundary terms)
\[
\mu_c = \frac{\delta F}{\delta \n} = \f - \d_{m_1} \f_{m_1} + \d_{m_1 m_2} \f_{m_1 m_2},
\]
which is Eq. (\ref{eq:mu_c_in_terms_of_f_hat})
in 2 dimensions with a free energy limited to second order gradients of $\n$. 
The same combinatorial factor appears when writing $\d_j \tilde f$ in terms of $\f_{m_1....m_\a}$ as a sum over all indices. 


\section{PFC mode expansions}

\subsection{2D hexagonal PFC}
\label{sec:appendix_hexagonal_mode_expansion}

We consider the 2D hexagonal PFC in the one-mode expansion for a macroscopic displacement field $\vec u$ at constant macroscopic density
\[
\psi = \psi_{eq}^{hex}(\vec r-\vec u) \equiv \P + \A,
\]
where 
\[
\A =  A_{hex}\sum_{\vec q_n \in \R_{hex}^{(1)}} e^{i\vec q_n \cdot (\vec r - \vec u)}
\equiv  A_{hex} \sum_{\vec q_n \in \R_{hex}^{(1)}} \fq.
\]
Since $\vec u$ varies slowly on the macroscopic scale, the resonance condition dictates 
\[
\avg \fq = 0,
\]
and 
\[
\avg{\fq \fqp} = \delta_{n,-n'},
\]
as we only get a non-zero average at resonance, when $\vec q_{n'} = -\vec q_{n}$ (depending on the method of coarse graining, this may be an approximation, albeit a very good one, see footnote \cite{footnote_coarse_grained_delta_function}).
Inserting the distorted PFC into the stress tensor, and using these identities, we get 
\[
\begin{split}
h_{ij}^{(2D~hex)} &= 
- 2\avg{(\L_1 \psi) \d_{ij} \psi},\\
&=
- 2\avg{(\L_1 \P) \d_{ij} \P}
- 2\avg{(\L_1 \A) \d_{ij} \A} \\
&=
- 2\avg{(\L_1 \A) \d_{ij} \A}
\end{split}
\]
First, we calculate, to first order in $\d_k u_l$, 
\[
\begin{split}
\L_1 \A 
&=
A_{hex}\sum_{\vec q_n \in \R_{hex}^{(1)}} \left (
1
-(q_{nk} - q_{nl} \d_k u_l)^2
\right )
\fq
\\
&=2 A_{hex} \d_k u_l
\sum_{\vec q_n \in \R_{hex}^{(1)}}
q_{nk} q_{nl} \fq,
\end{split}
\]

from which we get
\[
\begin{split}
\avg{(\L_1 \A) \d_{ij} \A}
&=
2 A_{hex}^2 \sum_{\vec q_n \in \R_{hex}^{(1)}}
\sum_{\vec q_{n'} \in \R_{hex}^{(1)}}
q_{nk} q_{nl}
(-q_{n'i} q_{n'j}) \avg{\fq \fqp} \\
&= -2 A_{hex}^2 
\sum_{\vec q_n \in \R_{hex}^{(1)}}
q_{ni} q_{nj} 
q_{nk} q_{nl},
\end{split}
\]
so 
\[
h_{ij}^{(2D~hex)} 
=
4 A_{hex}^2 
\sum_{\vec q_n \in \R_{hex}^{(1)}}
q_{ni} q_{nj} 
q_{nk} q_{nl}.
\]


\subsection{2D square PFC}
\label{sec:appendix_square_mode_expansion}

We consider the 2D square PFC in the two-mode expansion for a macroscopic displacement field $\vec u$ at constant macroscopic density
\[
\psi = \P + \A + \B,
\]
where
\[
\A 
= A_{sq} \sum_{\vec q_n \in \R_{sq}^{(1)}} e^{i\vec q_n \cdot (\vec r - \vec u)}
\equiv A_{sq} \sum_{\vec q_n \in \R_{sq}^{(1)}} \fq,
\]
\[
\B = B_{sq} \sum_{\vec p_n \in \R_{sq}^{(2)}} e^{i\vec p_n \cdot (\vec r - \vec u)}
\equiv B_{sq} \sum_{\vec p_n \in \R_{sq}^{(2)}} \fp.
\]
By resonance conditions, we have 
\[
\avg \fq = \avg \fp = \avg{\fq \fpp} = 0,
\]
and 
\[
\avg{\fq \fqp} = \avg{\fp \fpp} = \delta_{n',-n},
\]
from which, we find 
\[\begin{split}
h_{ij}^{(2D~sq)}
&=
-2 \avg{(\L_1 \L_2 \psi)(\L_1 + \L_2) \d_{ij} \psi }  \\
&=
-2 \avg{(\L_1 \L_2 \P)(\L_1 + \L_2) \d_{ij} \P } 
-2 \avg{(\L_1 \L_2 \A)(\L_1 + \L_2) \d_{ij} \A } 
-2 \avg{(\L_1 \L_2 \B)(\L_1 + \L_2) \d_{ij} \B } \\
&=
-2 \avg{(\L_1 \L_2 \A)(\L_1 + \L_2) \d_{ij} \A } 
-2 \avg{(\L_1 \L_2 \B)(\L_1 + \L_2) \d_{ij} \B } .
\end{split}
\]
To first order in $\d_k u_l$, we have
\[
\begin{split}
\L_1 \L_2 \A
&=
\L_2 \L_1 \A \\
&=
\L_2 \left (2 A_{sq} \d_k u_l \sum_{\vec q_n \in \R_{sq}^{(1)}} q_{nk} q_{nl} \fq \right )
 \\
&= 
2 A_{sq} \d_k u_l \sum_{\vec q_n \in \R_{sq}^{(1)}} q_{nk} q_{nl} (2-\vec q_{n'}^2) \fq\\
&=2 A_{sq} \d_k u_l \sum_{\vec q_n \in \R_{sq}^{(1)}} q_{nk} q_{nl}  \fq,
\end{split}
\]
so
\[
\begin{split}
\avg{(\L_1 \L_2 \A)(\L_1 + \L_2) \d_{ij} \A } 
&=
\avg{(\L_1 \L_2 \A)\L_2 \d_{ij} \A } 
\\
&=
2 A_{sq}^2 \d_k u_l \sum_{\vec q_n \in \R_{sq}^{(1)}}\sum_{\vec q_{n'} \in \R_{sq}^{(1)}} q_{nk} q_{nl} (2-\vec q_{n'}^2) (-q_{n'i} q_{n'j}) \avg{\fq \fqp}
\\
&=
-2 A_{sq}^2 \d_k u_l \sum_{\vec q_n \in \R_{sq}^{(1)}} q_{nk} q_{nl} q_{ni} q_{nj},
\end{split}
\]
and
\[
\begin{split}
\L_1 \L_2 \B &= 
\L_1 \left (
2 B_{sq} \d_k u_l \sum_{\vec p_n \in \R_{sq}^{(2)}} p_{nk} p_{nl} \fp
\right )\\
&=2 B_{sq} \d_k u_l \sum_{\vec p_n \in \R_{sq}^{(2)}} p_{nk} p_{nl} (1-\vec p_{n'}^2) \fp \\
&= - 2 B_{sq} \d_k u_l \sum_{\vec p_n \in \R_{sq}^{(2)}} p_{nk} p_{nl} \fp,
\end{split}
\]
so
\[
\begin{split}
\avg{(\L_1 \L_2 \B)(\L_1 + \L_2) \d_{ij} \B } 
&=
\avg{(\L_1 \L_2 \B)\L_1 \d_{ij} \B } \\
&= - 2 B_{sq}^2 \d_k u_l \sum_{\vec p_n \in \R_{sq}^{(2)}} \sum_{\vec p_{n'} \in \R_{sq}^{(2)}} p_{nk} p_{nl} (1-\vec p_{n'}^2) (-p_{n'i} p_{n'j}) \avg {\fp \fpp} ,\\
&=
- 2 B_{sq}^2 \d_k u_l
\sum_{\vec p_n \in \R_{sq}^{(2)}} p_{ni} p_{nj} p_{nk} p_{nl},
\end{split}
\]
which gives
\[
h_{ij}^{(2D~sq)} 
=
4 A_{sq}^2 \d_k u_l \sum_{\vec q_n \in \R_{sq}^{(1)}} q_{nk} q_{nl} q_{ni} q_{nj}
+4 B_{sq}^2 \d_k u_l
\sum_{\vec p_n \in \R_{sq}^{(2)}} p_{ni} p_{nj} p_{nk} p_{nl}.
\]


\subsection{3D bcc PFC}
\label{sec:appendix_bcc_mode_expansion}

We consider the bcc PFC in the one-mode expansion for a macroscopic displacement field $\vec u$ at constant macroscopic density
\[
\psi = \psi_{eq}^{bcc}(\vec r-\vec u) \equiv \P + \A,
\]
where 
\[
\A =  A_{bcc}\sum_{\vec q_n \in \R_{bcc}^{(1)}} e^{i\vec q_n \cdot (\vec r - \vec u)}
\equiv  A_{bcc} \sum_{\vec q_n \in \R_{bcc}^{(1)}} \fq.
\]
By resonance conditions, we have
\[
\avg \fq = 0,
\]
 and
\[
\avg{\fq \fqp} = \delta_{n,-n'}
\]
as we only get a non-zero average at resonance, when $\vec q_{n'} = -\vec q_{n}$.
Inserting the distorted PFC into the stress tensor, and using these identities, we get 
\[
\begin{split}
h_{ij}^{(3D~bcc)} &= 
- 2\avg{(\L_1 \psi) \d_{ij} \psi},\\
&=
- 2\avg{(\L_1 \P) \d_{ij} \P}
- 2\avg{(\L_1 \A) \d_{ij} \A} \\
&=
- 2\avg{(\L_1 \A) \d_{ij} \A}.
\end{split}
\]
First, we calculate, to first order in $\d_k u_l$, 
\[
\begin{split}
\L_1 \A 
&=
A_{bcc}\sum_{\vec q_n \in \R_{bcc}^{(1)}} \left (
1
-(q_{nk} - q_{nl} \d_k u_l)^2
\right )
\fq
\\
&=2 A_{bcc} \d_k u_l
\sum_{\vec q_n \in \R_{bcc}^{(1)}}
q_{nk} q_{nl} \fq,
\end{split}
\]
from which we get
\[
\begin{split}
\avg{(\L_1 \A) \d_{ij} \A}
&=
2 A_{bcc}^2 \sum_{\vec q_n \in \R_{bcc}^{(1)}}
\sum_{\vec q_{n'} \in \R_{bcc}^{(1)}}
q_{nk} q_{nl}
(-q_{n'i} q_{n'j}) \avg{\fq \fqp} \\
&= -2 A_{bcc}^2 
\sum_{\vec q_n \in \R_{bcc}^{(1)}}
q_{ni} q_{nj} 
q_{nk} q_{nl},
\end{split}
\]
so 
\[
h_{ij}^{(3D~bcc)} 
=
4 A_{bcc}^2 
\sum_{\vec q_n \in \R_{bcc}^{(1)}}
q_{ni} q_{nj} 
q_{nk} q_{nl}.
\]


\subsection{3D fcc PFC}
\label{sec:appendix_fcc_mode_expansion}

We consider the fcc PFC in the two-mode expansion for a macroscopic displacement field $\vec u$ at constant macroscopic density
\[
\psi = \P + \A + \B,
\]
where
\[
\A 
= A_{fcc} \sum_{\vec q_n \in \R_{fcc}^{(1)}} e^{i\vec q_n \cdot (\vec r - \vec u)}
\equiv A_{fcc} \sum_{\vec q_n \in \R_{fcc}^{(1)}} \fq,
\]
\[
\B = B_{fcc} \sum_{\vec p_n \in \R_{fcc}^{(4/3)}} e^{i\vec p_n \cdot (\vec r - \vec u)}
\equiv B_{fcc} \sum_{\vec p_n \in \R_{fcc}^{(4/3)}} \fp.
\]
By resonance conditions, we have 
\[
\avg \fq = \avg \fp = \avg{\fq \fpp} = 0,
\]
and 
\[
\avg{\fq \fqp} = \avg{\fp \fpp} = \delta_{n',-n},
\]
from which, we find 
\[\begin{split}
h_{ij}^{(3D~fcc)}
&=
-2 \avg{(\L_1 \L_{\frac 4 3} \psi)(\L_1 + \L_{\frac 4 3}) \d_{ij} \psi }  \\
&=
-2 \avg{(\L_1 \L_{\frac 4 3} \P)(\L_1 + \L_{\frac 4 3}) \d_{ij} \P } 
-2 \avg{(\L_1 \L_{\frac 4 3} \A)(\L_1 + \L_{\frac 4 3}) \d_{ij} \A } 
-2 \avg{(\L_1 \L_{\frac 4 3} \B)(\L_1 + \L_{\frac 4 3}) \d_{ij} \B } \\
&=
-2 \avg{(\L_1 \L_{\frac 4 3} \A)(\L_1 + \L_{\frac 4 3}) \d_{ij} \A } 
-2 \avg{(\L_1 \L_{\frac 4 3} \B)(\L_1 + \L_{\frac 4 3}) \d_{ij} \B }.
\end{split}
\]
To first order in $\d_k u_l$, we have
\[
\begin{split}
\L_1 \L_{\frac 4 3} \A
&=
\L_{\frac 4 3} \L_1 \A \\
&=
\L_{\frac 4 3} \left (2 A_{fcc} \d_k u_l \sum_{\vec q_n \in \R_{fcc}^{(1)}} q_{nk} q_{nl} \fq \right )
 \\
&= 
2 A_{fcc} \d_k u_l \sum_{\vec q_n \in \R_{fcc}^{(1)}} q_{nk} q_{nl} (\frac{4}{3}-\vec q_{n'}^2) \fq\\
&=\frac{2}{3} A_{fcc} \d_k u_l \sum_{\vec q_n \in \R_{fcc}^{(1)}} q_{nk} q_{nl}  \fq,
\end{split}
\]
so
\[
\begin{split}
\avg{(\L_1 \L_{\frac 4 3} \A)(\L_1 + \L_{\frac 4 3}) \d_{ij} \A } 
&=
\avg{(\L_1 \L_{\frac 4 3} \A)\L_{\frac 4 3} \d_{ij} \A } 
\\
&=
\frac{2}{3} A_{fcc}^2 \d_k u_l \sum_{\vec q_n \in \R_{fcc}^{(1)}}\sum_{\vec q_{n'} \in \R_{fcc}^{(1)}} q_{nk} q_{nl} (\frac{4}{3}-\vec q_{n'}^2) (-q_{n'i} q_{n'j}) \avg{\fq \fqp}, 
\\
&=
-\frac{2}{9} A_{fcc}^2 \d_k u_l \sum_{\vec q_n \in \R_{fcc}^{(1)}} q_{nk} q_{nl} q_{ni} q_{nj},
\end{split}
\]
and
\[
\begin{split}
\L_1 \L_{\frac 4 3} \B &= 
\L_1 \left (
2 B_{fcc} \d_k u_l \sum_{\vec p_n \in \R_{fcc}^{(4/3)}} p_{nk} p_{nl} \fp
\right )\\
&=2 B_{fcc} \d_k u_l \sum_{\vec p_n \in \R_{fcc}^{(4/3)}} p_{nk} p_{nl} (1-\vec p_{n'}^2) \fp \\
&= - \frac{2}{3} B_{fcc} \d_k u_l \sum_{\vec p_n \in \R_{fcc}^{(4/3)}} p_{nk} p_{nl} \fp,
\end{split}
\]
so
\[
\begin{split}
\avg{(\L_1 \L_{\frac 4 3} \B)(\L_1 + \L_{\frac 4 3}) \d_{ij} \B } 
&=
\avg{(\L_1 \L_2 \B)\L_1 \d_{ij} \B } \\
&= - \frac{2}{3} B_{fcc}^2 \d_k u_l \sum_{\vec p_n \in \R_{fcc}^{(4/3)}} \sum_{\vec p_{n'} \in \R_{fcc}^{(4/3)}} p_{nk} p_{nl} (1-\vec p_{n'}^2) (-p_{n'i} p_{n'j}) \avg {\fp \fpp} ,\\
&=
-\frac{2}{9} B_{fcc}^2 \d_k u_l
\sum_{\vec p_n \in \R_{fcc}^{(4/3)}} p_{ni} p_{nj} p_{nk} p_{nl},
\end{split}
\]
which gives
\[
h_{ij}^{(3D~fcc)} 
=
\frac{4}{9} A_{fcc}^2 \d_k u_l \sum_{\vec q_n \in \R_{fcc}^{(1)}} q_{nk} q_{nl} q_{ni} q_{nj}
+\frac{4}{9} B_{fcc}^2 \d_k u_l
\sum_{\vec p_n \in \R_{fcc}^{(4/3)}} p_{ni} p_{nj} p_{nk} p_{nl}.
\]


\subsection{3D sc PFC}
\label{sec:appendix_sc_mode_expansion}

We consider the sc PFC in the three-mode expansion for a macroscopic displacement field $\vec u$ at constant macroscopic density
\[
\psi  = \P + \A + \B + \C,
\]
where
\[
\A = A_{sc} \sum_{\vec q_n \in \R_{sc}^{(1)}} \fq,
\]
\[
\B = B_{sc} \sum_{\vec q_n \in \R_{sc}^{(2)}} \fp,
\]
\[
\C = C_{sc} \sum_{\vec t_n \in \R_{sc}^{(3)}} \ft,
\]
By resonance conditions, we have 
\[\avg \fq = \avg \fp = \avg{\ft} = \avg{\fp \ftp} =\avg{\fq \ftp} = \avg{\fq \fpp} = 0,
\] 
and 
\[
\avg{\fq \fqp} = \avg{\fp \fpp} = \avg{\ft \ftp} =\delta_{n',-n},
\]
From which we get 
\[\begin{split}
h_{ij}^{(3D~sc)}
&=
-2\avg{(\L_1 \L_2 \L_3 \psi)(\L_2 \L_3 + \L_1 \L_3 + \L_1 \L_2) \d_{ij} \psi} \\
&=
-2\avg{(\L_1 \L_2 \L_3 \P)(\L_2 \L_3 + \L_1 \L_3 + \L_1 \L_2) \d_{ij} \P}
-2\avg{(\L_1 \L_2 \L_3 \A)(\L_2 \L_3 + \L_1 \L_3 + \L_1 \L_2) \d_{ij} \A}\\
&
\hspace{0.3cm}-2\avg{(\L_1 \L_2 \L_3 \B)(\L_2 \L_3 + \L_1 \L_3 + \L_1 \L_2) \d_{ij} \B}
-2\avg{(\L_1 \L_2 \L_3 \C)(\L_2 \L_3 + \L_1 \L_3 + \L_1 \L_2) \d_{ij} \C}\\
&=
-2\avg{(\L_1 \L_2 \L_3 \A)(\L_2 \L_3 + \L_1 \L_3 + \L_1 \L_2) \d_{ij} \A
-2\avg{(\L_1 \L_2 \L_3 \B)(\L_2 \L_3 + \L_1 \L_3 + \L_1 \L_2) \d_{ij} \B}}\\
&
\hspace{7.5cm}
-2\avg{(\L_1 \L_2 \L_3 \C)(\L_2 \L_3 + \L_1 \L_3 + \L_1 \L_2) \d_{ij} \C}.
\end{split}\]
To first order in $\d_k u_l$, we have
\[\begin{split}
\L_1 \L_2 \L_3 \A
&=
\L_2 \L_3 \L_1 \A \\
&=
2A_{sc}\L_2 \L_3 \d_k u_l \sum_{\vec q_n \in \R_{sc}^{(1)}} q_{nk} q_{nl} \fq\\
&=
2A_{sc} \d_k u_l  \sum_{\vec q_n \in \R_{sc}^{(1)}} (2-\vec q_n^2)(3-\vec q_n^2) q_{nk} q_{nl} \fq\\
&= 4 A_{sc}\d_k u_l \sum_{\vec q_n \in \R_{sc}^{(1)}} q_{nk} q_{nl} \fq,
\end{split}\]
so
\[\begin{split}
\avg{(\L_1 \L_2 \L_3 \A)(\L_2 \L_3 + \L_1 \L_3 + \L_1 \L_2) \d_{ij} \A}
&=\avg{(\L_1 \L_2 \L_3 \A)\L_2 \L_3  \d_{ij} \A}
\\
&= 
 4 A_{sc}^2\d_k u_l \sum_{\vec q_n \in \R_{sc}^{(1)}} q_{nk} q_{nl} (- q_{n'i} q_{n'j}) 
 (2-\vec q_{n'}^2)(3-\vec q_{n'}^2) 
 \avg{\fq \fqp}\\
&=
-8 A_{sc}^2 \d_k u_l
\sum_{\vec q_n \in \R_{sc}^{(1)}}
q_{ni} q_{nj} q_{nk} q_{nl},
\end{split}\]
\[\begin{split}
    \L_1 \L_2 \L_3 \B 
    &=
    \L_1 \L_3 \L_2 \B \\
    &=
    2 B_{sc} \d_k u_l \L_1 \L_3 \sum_{\vec p_n \in \R_{sc}^{(2)}} p_{nk} p_{nl} \fp\\
    &=
    2 B_{sc} \d_k u_l  \sum_{\vec p_n \in \R_{sc}^{(2)}} (1-\vec p_n^2)(3-\vec p_n^2) p_{nk} p_{nl} \fp\\
    &=
    -2B_{sc} \d_k u_l \sum_{\vec p_n \in \R_{sc}^{(2)}} p_{nk} p_{nl} \fp,
\end{split}\]
so
\[\begin{split}
\avg{(\L_1 \L_2 \L_3 \B)(\L_2 \L_3 + \L_1 \L_3 + \L_1 \L_2) \d_{ij} \B}
&=\avg{(\L_1 \L_2 \L_3 \B)\L_1 \L_3 \d_{ij} \B}
\\
&= 
 -2 B_{sc}^2\d_k u_l \sum_{\vec p_n \in \R_{sc}^{(2)}} p_{nk} p_{nl} (- p_{n'i} p_{n'j}) 
 (1-\vec p_{n'}^2)(3-\vec p_{n'}^2) 
 \avg{\fp \fpp}\\
&=
-2 B_{sc}^2 \d_k u_l
\sum_{\vec p_n \in \R_{sc}^{(2)}}
p_{ni} p_{nj} p_{nk} p_{nl},
\end{split}\]
and
\[\begin{split}
    \L_1 \L_2 \L_3 \C 
    &=
    2 C_{sc} \d_k u_l \L_1 \L_2 \sum_{\vec t_n \in \R_{sc}^{(3)}} t_{nk} t_{nl} \ft\\
    &=
    2 C_{sc} \d_k u_l  \sum_{\vec t_n \in \R_{sc}^{(3)}} (1-\vec t_n^2)(2-\vec t_n^2) t_{nk} t_{nl} \ft\\
    &=
    4C_{sc} \d_k u_l \sum_{\vec t_n \in \R_{sc}^{(3)}} p_{nk} p_{nl} \ft,
\end{split}\]
so
\[\begin{split}
\avg{(\L_1 \L_2 \L_3 \C)(\L_2 \L_3 + \L_1 \L_3 + \L_1 \L_2) \d_{ij} \C}
& \avg{(\L_1 \L_2 \L_3 \C)\L_1 \L_2 \d_{ij} \C}
\\
&= 
 4 C_{sc}^2 \d_k u_l \sum_{\vec t_n \in \R_{sc}^{(3)}} t_{nk} t_{nl} (- t_{n'i} t_{n'j}) 
 (1-\vec t_{n'}^2)(2-\vec t_{n'}^2)
 \avg{\ft \ftp}\\
&=
-8 C_{sc}^2 \d_k u_l
\sum_{\vec t_n \in \R_{sc}^{(3)}}
t_{ni} t_{nj} t_{nk} t_{nl}.
\end{split}\]
Thus, we find
\[
h_{ij}^{(3D~sc)}
=
16 A_{sc}^2 \d_k u_l
\sum_{\vec q_n \in \R_{sc}^{(1)}}
q_{ni} q_{nj} q_{nk} q_{nl}
+
4 B_{sc}^2 \d_k u_l
\sum_{\vec p_n \in \R_{sc}^{(2)}}
p_{ni} p_{nj} p_{nk} p_{nl}
+
16 C_{sc}^2 \d_k u_l
\sum_{\vec t_n \in \R_{sc}^{(3)}}
t_{ni} t_{nj} t_{nk} t_{nl}.
\]

\end{widetext}

\bibliography{references}

\begin{thebibliography}{10}
\bibitem{footnote_Einstein_summation_convention}
Here, and throughout this article, we use the summation convention for repeated
  indices.
\bibitem{footnote_coarse_grained_delta_function}
Here, we have implicitly imposed a convention of the numbering $n \in \mathbb
  Z$ of reciprocal lattice vectors. Since reciprocal lattice vectors come in
  oppositely signed pairs, we are free to define an indexing in which $\vec
  K_{-n} = -\vec K_n$ for all $n$. This convention sets $\vec K_0 = \vec 0$.
  Additionally, we would only obtain an exact delta function if coarse graining
  was performed over an exact number of unit cells. With the coarse graining
  definition of Eq. (\ref{eq:coarse_graining_definition}), the value of
  $\avg{e^{i(\vec K_n + \vec K_{n'}) \cdot \vec r}}$ will be $e^{-\frac{1}{2}
  a_0^2 (\vec K_n + \vec K_{n'})^2}$, which is $1$ for $n=-n'$ and for a vector
  sum $(\vec K_n + \vec K_{n'})$ giving one of the closest reciprocal modes,
  where $ a_0 |\vec K_n + \vec K_{n'}| \sim 2\pi$, we get $e^{-\frac{1}{2}
  a_0^2 (\vec K_n + \vec K_{n'})^2} \sim e^{-2\pi^2}=2.67\times 10^{-9}$.
\end{thebibliography}

\end{document}